\documentclass[aps,pra,amsmath,amssymb,groupedaddress,10pt,superscriptaddress,floatfix,twocolumn,showkeys]{revtex4-1}
\usepackage[T1]{fontenc}
\usepackage{color}
\usepackage{amsmath}
\usepackage{siunitx}
\usepackage{xfrac}
\usepackage{braket}
\usepackage[mathscr]{euscript}

\newcommand{\tr}{\text{tr}}
\newcommand{\scL}{\mathscr{L}}
\newcommand{\scR}{\mathscr{R}}
\newcommand{\scS}{\mathscr{S}}

\usepackage[normalem]{ulem}

\def\tr{\mbox{tr}}
\def\eq{\begin{eqnarray}}
\def\en{\end{eqnarray}}

\def\beq{\begin{eqnarray}}
\def\een{\end{eqnarray}}
\def\bfig{\begin{figure}}
\def\efig{\end{figure}}

\begin{document}

\title{The Hong-Ou-Mandel effect with atoms}

\author{A. M. Kaufman}
\affiliation{JILA, National Institute of Standards and Technology and University of Colorado, and
Department of Physics, University of Colorado, Boulder, Colorado 80309, USA}
\author{M. C. Tichy}
\affiliation{Department of Physics and Astronomy, University of Aarhus, DK--8000 Aarhus C, Denmark}
\author{F. Mintert}
\affiliation{Department of Physics, Imperial College, SW7 2AZ London, UK}
\author{A. M. Rey}
\affiliation{JILA, National Institute of Standards and Technology and University of Colorado, and
Department of Physics, University of Colorado, Boulder, Colorado 80309, USA}
\author{C. A. Regal}
\affiliation{JILA, National Institute of Standards and Technology and University of Colorado, and
Department of Physics, University of Colorado, Boulder, Colorado 80309, USA}

\begin{abstract}

Controlling light at the level of individual photons has led to advances in fields ranging from quantum information and precision sensing to fundamental tests of quantum mechanics. A central development that followed the advent of single photon sources was the observation of the Hong-Ou-Mandel (HOM) effect, a novel two-photon path interference phenomenon experienced  by indistinguishable  photons. The effect is now a central technique in the field of quantum optics, harnessed for a variety of applications such as diagnosing single photon sources and creating probabilistic entanglement in linear quantum computing.  Recently, several distinct experiments using atomic sources have realized the requisite control to observe and exploit Hong-Ou-Mandel interference of atoms. This article  provides a summary of this phenomenon and discusses some of its  implications for atomic systems. Transitioning from the domain of photons to atoms opens new perspectives on fundamental concepts, such as the classification of entanglement of identical particles. It aids in the design  of novel probes of quantities such as entanglement entropy by combining   well established  tools of AMO physics --- unity single-atom detection, tunable interactions, and scalability --- with the Hong-Ou-Mandel interference. Furthermore, it is now possible for established protocols in the photon community, such as measurement-induced entanglement, to be employed in atomic experiments that possess deterministic single-particle production and detection. Hence, the realization of the HOM effect with atoms represents a productive union of central ideas in quantum control of atoms and photons.

\end{abstract}

\keywords{Hong-Ou-Mandel effect, Entanglement, Ultracold atoms, Quantum statistics, Many-particle interference, Entanglement entropy, Schmidt rank, Bose-Hubbard model}

\maketitle

\tableofcontents


\section{Introduction}

\begin{figure}[h]
\includegraphics[width=\linewidth]{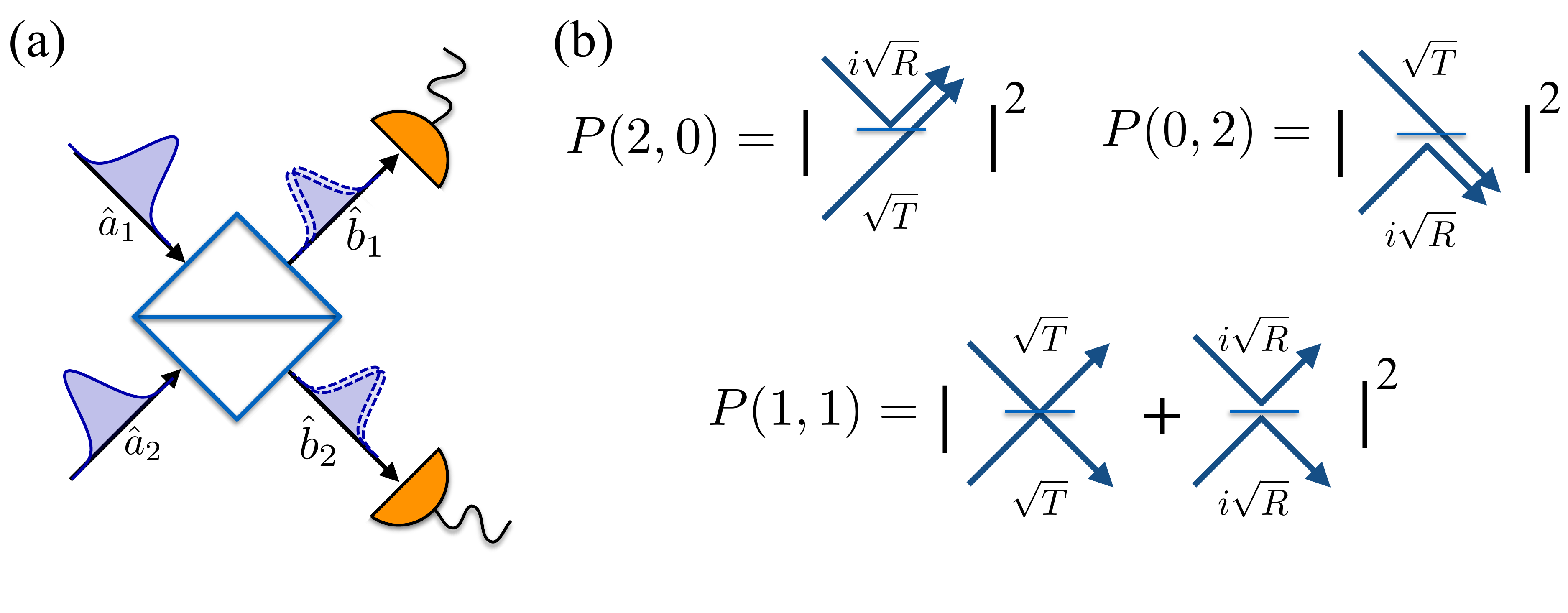}
\caption{Two-particle interference in the Hong-Ou-Mandel experiment. (a) Physical setup: Two identical particles impinge on the two input modes $(\hat{a}_1,\hat{a}_2)$ of a beam-splitter and emerge in two output modes $(\hat{b}_1,\hat{b}_2)$. (b) Evaluation of event probabilities via Feynman paths. In the HOM effect the bosonic nature of photons results in destructive interference of the P(1,1) outcome. (c) Mandel dip as a signature of the HOM effect in a photon experiment.  The distinguishability of the photons is tuned by translating the beamsplitter to give a relative photon delay; at zero delay $P(1,1)$ dips to 0.  } \label{HOMSketch}
\end{figure}

The Hong-Ou-Mandel (HOM) effect \cite{HOM} is a two-particle interference phenomenon, wherein two indistinguishable bosonic particles are interfered on a beamsplitter, and the particles always emerge on the same, but random output port (Fig.~\ref{HOMSketch}).  In the original HOM experiment, two single photons were sent into each port of a balanced beamsplitter, and the population of the output ports was measured with single-photon detectors. The experimenters quantified the probability for joint detection at the two detectors, that is, the likelihood that both detectors click within a small time window~\cite{HOM}.  Classically, one might compare the experiment to flipping two independent coins, where each coin represents a photon and the coin faces represent the output port on which the photon emerges. Quantum mechanics dictates that  photons indistinguishable in all degrees of freedom besides the initial mode they occupy will always emerge on the same output port, akin to the two independent coins always landing identically.  Experimentally, the quantum mechanical outcome leads to the ``Mandel dip'' in the joint detection probability, which ideally extends to zero as the photons are tuned to impinge on the beamsplitter at exactly the same time. The effect is traced to the bosonic nature of the photons, and a resulting destructive interference of the paths that yield joint detection. If photons were fermionic, Pauli exclusion would prevent the particles from being in the same mode and lead to a preponderance of joint detections at the output ports.

The HOM effect plays an important role in quantum optics and quantum information. It has become one of the textbook examples of an observed quantum interference phenomenon that cannot be explained by semiclassical theory.  Further, it is a standard technique for characterizing single-photon sources and is the key component of linear optical quantum computation protocols~\cite{Knill2001}.  Generalization of the HOM effect to many-particle and many-mode interference leads to computationally complex situations such as the boson sampling problem~\cite{Knill2001,Aaronson:2010fk}.  In recent years, a number of systems beyond optical photons have come to observe Hong-Ou-Mandel physics.  Microwave-photons and phonons in ion traps have demonstrated the HOM effect~\cite{Walraff2013,Toyoda2015}; electrons have realized the fermionic counterpart to HOM~\cite{electron}; and, Rydberg interactions in ensembles have been used to realize HOM interference of collective excitations~\cite{Li2016} . Further, as is a focus of this article, atoms can serve  as ``single particle sources'' in analogy to ``single photon sources'' used in the  HOM effect. 

The challenge in realizing the HOM effect with atoms is placing the particles in equivalent modes, defined by both motion and spin, and realizing effective atomic beamsplitters.  In one paradigm, atoms are held in separate potential wells, such as in an optical lattice or a double-well formed by focused laser beams known as optical tweezers.  Tunneling between the potential wells then acts as an effective beamsplitter.
Pairs of trapped atoms  in the required pure single-particle states can be prepared independently, as for example in experiments in which individual $^{87}$Rb atoms were laser-cooled to their motional ground state~\cite{Kaufman2014}.  Alternatively, the atoms can be prepared collectively, as in optical lattice experiments in which a Mott insulator is created via a phase transition from a superfluid and observed with a quantum-gas microscope~\cite{Bakr2011,Preiss2015}. A second paradigm uses freely-propagating atom sources.  Experiments utilizing ultracold $^4$He atoms have created twin beams that can be interfered with tunable temporal overlap, in analogy to the original HOM photonic beamsplitter~\cite{Lopes2015}.

In addition to state preparation of individual atoms, the observation of the HOM effect with atoms requires achieving negligible atom-atom interactions by appropriately tuning the atomic scattering length or confinement. Manipulating non-interacting atoms not only allows the  exploration of the HOM effect, but opens an entire paradigm for creating entangled quantum states that use measurement instead of interaction.  These ideas have long been explored in the context of linear optics with single photons. Such experiments feature a surprisingly rich space that can result from combining projective measurement with only linear elements such as beamsplitters and waveplates~\cite{Knill2001}.  On the other hand, trapped atoms offer controllability that may enable explorations that are difficult with photons -- for example exploration of the role of quantum statistics on distant atoms through interchange of two particles~\cite{Roos2017}.

We have multi-fold goals for this article. We would like to provide a background on optical HOM experiments for the atomic physics community; there are interesting parallels between the HOM effect experiments that have now been performed with atoms and those with photons.  For a rounded understanding of the HOM effect, it is useful to realize that the HOM effect incorporates both coherent superposition of paths (modes) with single-photon inputs and the bosonic nature of photons~\cite{TichyThesis}, which demands proper
 wavefunction symmetrization. As such, the HOM effect is an excellent thought experiment for understanding entanglement in the presence of symmetrization. One potential misconception is to assume that symmetrization automatically implies entanglement due to the non-separability of the symmetrized wavefunction.  We review one formalism that helps clarify the apparent contradiction, namely, the presence of non-separability but the absence of the quantum fluctuations fundamental to entanglement~\cite{ItalianPRA,TichyJOP,Cirac2012,TichyThesis,Barnum2004,Paunkovic2004}.

Further, the HOM effect is only the simplest case of richer types of many-particle interference, in which more particles and more modes  participate. Those have been the subject of considerable studies in the optics community, both theoretically \cite{Lim:2005qt,Tichy:2012NJP,PhysRevA.91.013811} and experimentally ~\cite{Ou:2007ly,Spagnolo:2013fk,Ra:2013kx,Jin2016,Huang2011,Yao2012,PhysRevLett.117.210502}. Atomic systems seem particularly well-suited to future studies of many-particle interference by combining the toolset required for two-particle quantum interference (the HOM effect) with the scalability and measurement techniques endemic to cold atomic sources.  While the topic of fully general multipartite entanglement characterization is beyond the scope of this article, we present a basic delineation between the kinds of entanglement present in two-particle quantum states of identical particles.

Lastly, while photons in vacuum are non-interacting, atoms of course can interact, and their interactions can play an important role in experiments.  It is thus useful to study how interactions affect many-particle interference,  the range of entangled states that can be created with interactions, and  how the  HOM effect can be used to characterize  the emergent  many-body states. Already experiments with ultracold atoms in lattices have taken the concept of HOM beyond two particles:  Interfering copies of a quantum state have  been used to quantify entanglement entropy of an interacting gas~\cite{Daley,Islam2015}, and to study quantum thermalization through entanglement~\cite{Kaufman2016}.

\section{Experiments}
\label{sec:experiment}

\subsection{Photon HOM experiments and photon sources}

The HOM effect, and its name, come from an experiment with single photons created via parametric downconversion in which a single pump photon is split into a pair of lower-frequency photons~\cite{HOM}.  The original experiment was titled ``Measurement of sub-picosecond time intervals between two photons via interference''.  The technique used quantum interference to determine if two photons impinged simultaneously upon the beamsplitters and hence were indistinguishable in their arrival times.  The photons used in HOM's experiment were so-called twin pairs of entangled photons.  If these pairs are permutation symmetric one sees the bosonic HOM-effect, but on the other hand if one uses bosons entangled in another degree of freedom with anti-symmetric spatial wave-functions, they will leave in different ports as if they were fermions ~\cite{Obrien2013,Keilmann2010}. One of the most important aspects of the  the Hong-Ou-Mandel effect is the fact that it is a quantum phenomenon that does not require the incident bosonic particles be correlated to begin with; indistinguishability is all that is required. The photons can originate completely independently and the HOM effect still persists.  Following the original HOM experiment, efforts were made to realize the effect with independent photon sources.  This was first explored in experiments of Ref.~\cite{Rarity1996} and after it  many subsequent experiments have been carried out~\cite{Riedmatten2003,Kaltenbach2006,Pittman1996}.  For one particularly elucidating example, the HOM effect has been observed by creating two independent photons via radiation decay from two separately trapped atoms~\cite{Beugnon2006}.

\begin{figure*}[]
\centering
\includegraphics[width=\linewidth]{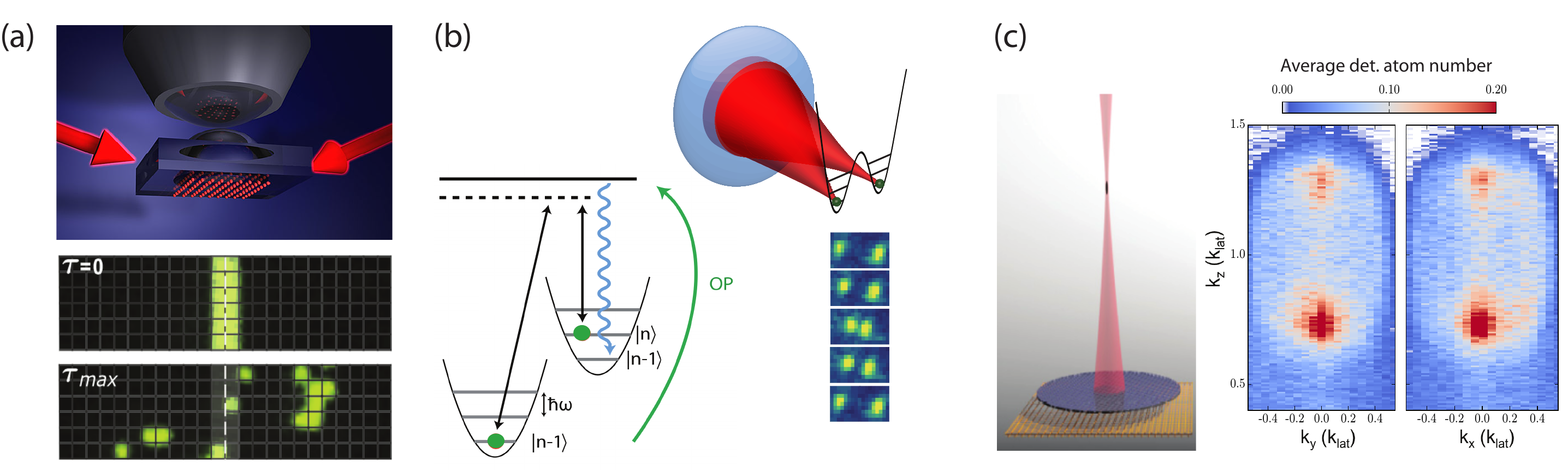}
\caption{Experimental realization of  single-atom sources analogous  to single-photon sources of quantum optics. Some examples of single-atom sources are:  (a) Atoms in optical lattices.  A microscope was used to individually observe atoms loaded from a Bose-Einstein condensate into a two dimensional lattice and to  image  their  quantum random walk trajectories (portions of figure taken from Ref.~\cite{Preiss2015}).  (b)  Atoms in optical tweezers. Individual neutral atoms were laser-cooled to their ground state~\cite{Kaufman2012} using alternating Raman sideband transitions and optical pumping (OP) between two hyperfine states. Optical tweezers were then used for real-time spatial positioning of the atoms (right images) and individual atom imaging.   (c) A twin-pair source of $^4$He. Twin-pairs were produced by a four-wave mixing process.  Multichannel detection (left) enabled spatio-temporal imaging and to probe the two-dimensional momentum distribution of the atom pairs (right). Images from Ref.~\cite{Lopesthesis}.} \label{exp}
\end{figure*}

\subsection{Single-atom sources}

Since the realization of Bose-Einstein condensation~\cite{Anderson1995}, ultra-cold atom experiments have become a useful laboratory for the investigation of the role of quantum statistics and indistinguishability in the behavior of many-particle systems. Atom optics phenomena such as the atom laser~\cite{Andrews1997,Ottl2005} and the Hanbury Brown and Twiss effect with gases of bosons were realized within a decade of Bose-Einstein condensates (BEC)~\cite{Jeltes2007,Dall2010,Schellekens2005,Perrin2012}.  Recently increasing interest has turned to single-atom imaging and control. In optical lattices, tunneling dynamics of individual atoms can be observed in quantum-gas microscopes~\cite{Bakr2011,BlochQGM}, and a variety of studies in which both interactions and quantum statistics play a role have been realized~\cite{Simon2011}.  In this article we will discuss different platforms that have  achieved the  single-atom control and pure state preparation required for realizing the HOM effect.  Those include laser-cooled optical tweezer arrays, optical lattices, and twin-pair sources from BECs. These systems and state preparation therein are summarized in Fig.~\ref{exp}.
In the following sections we discuss how the HOM effect has been observed in these different physical platforms.

\subsection{HOM effect with tunneling atoms}
\label{sec:double_well}

\begin{figure*}[t]
\includegraphics[width=0.9\linewidth]{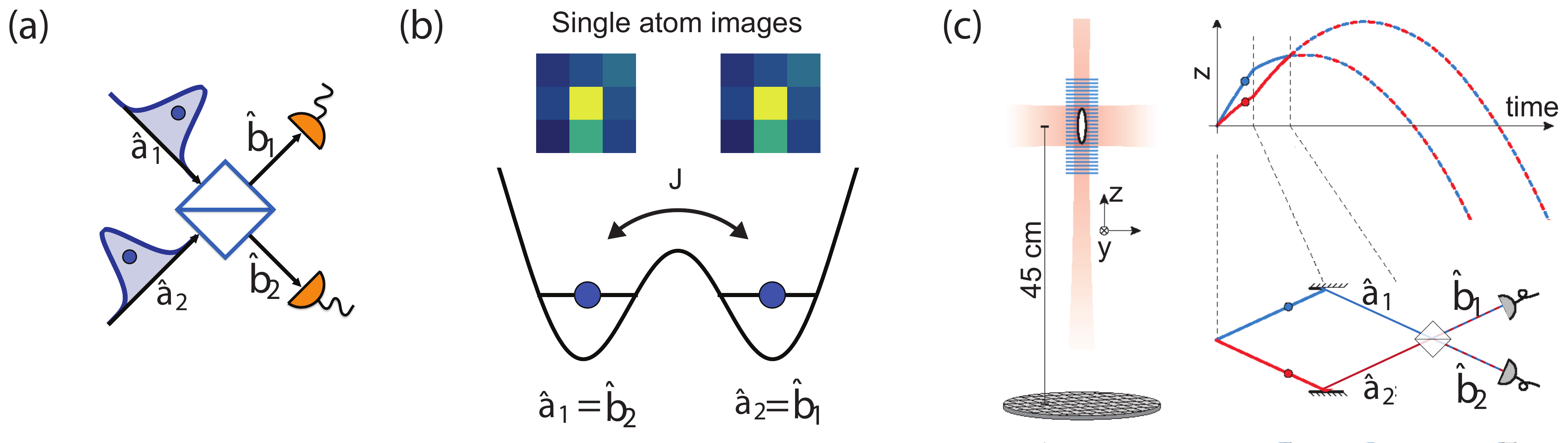}
\caption{Physical realizations of atomic beamsplitters for the HOM effect. (a) Free-space photonic beamsplitter with photons impinging on both input ports $\hat{a}_1$ and $\hat{a}_2$, and detected on two output ports $\hat{b}_1$ and $\hat{b_2}$.  (b) Double-well potential with tunneling, as can be created with optical tweezers or lattices.  Single atoms are mixed between the left and right wells by tunneling $J$, and detected at the output by parity imaging on each site. (c) Propagating twin-pairs are generated from a BEC source, reflected, and then beam-split through Bragg processes (Image adapted from Ref.~\cite{Lopes2015}).}
\label{fig:beamsplitter}
\end{figure*}

An atomic beamsplitter, as required for the HOM effect with atoms, can be realized by tunneling between closely-spaced potential wells, such as the double-well shown in Fig.~\ref{fig:beamsplitter}(b), which forms the minimal setup for observing the HOM effect in a bound system. Due to the tunnel-coupling, the energy-eigenstates are coherent superpositions of the ground states of each potential well $\ket{\scL}$ and $\ket{\scR}$. An initially prepared  atom  in the left well evolves into  a coherent superposition  of the form $\ket{\scL} \rightarrow \cos(Jt) \ket{\scL}  + i\sin(Jt) \ket{\scR}$, where $J$ is the tunneling rate.  Stopping the evolution at a specific time $t_{HOM}=\pi/4J$ yields an equal superposition associated with an effective $50-50$ beamsplitter ($R=T=1/2$).  Position-resolved single-atom detection then plays the role of photon detection at output ports of the beamsplitter.

In Ref.~\cite{Kaufman2014}, two optical tweezers were used to create a double-well potential in which a single atom was initialized on each well.  The experiment employed ground-state laser cooling in order to control all degrees of freedom of two independently prepared $^{87}$Rb atoms~\cite{Kaufman2012,Kaufmanthesis}. The trapping parameters were set to realize a sufficiently small ratio of interaction to tunneling and the atomic spin used as a knob to tune distinguishability.  In analogy to polarization of photons,  a Mandel dip was observed via rotation of the relative spin state~\cite{Kaufman2014}. The observation of the HOM effect in the experiment  demonstrated  the capability to create indistinguishable, and hence pure atomic quantum states in both motion and spin, and  showed  that  quantum statistics can strongly determine the interference properties of laser cooled atoms individually placed in their motional ground-state.

Experiments in optical lattices offer another way to initialize single bosons for observing quantum interference of single atoms. In fact, the use of optical lattices for the observation of the HOM effect has long been considered~\cite{Strabley,MeschedePNAS,Robens2016}. In recent successful efforts a Mott-insulator  transition \cite{Greiner2002} was used to create a uniformly-filled array of bosons.  Via lattice manipulations and quantum-gas microscopy single-atoms were interfered, and the particle number  measured on individual sites after interference. The interference protocol consisted in decreasing the lattice depth to allow particles to tunnel and thus to perform a random walk. In the regime where interactions were made smaller than tunneling, interference phenomena arising from quantum statistics and closely-related to HOM were observed~\cite{Preiss2015}. Quite recently, a generalized form of HOM was employed to study entanglement in ground state and non-equilibrium Bose-Hubbard systems~\cite{Islam2015,Kaufman2016}.

\subsection{HOM effect with twin-pair source}

Early ultracold-atom experiments probed density-density correlations in momentum distributions of expanding Bose or Fermi gases of alkali atoms and analyzed them using similar methods to typical photon-photon correlation experiments~\cite{Altman2004,Greiner2005,Folling2005}.  Experiments using metastable helium atoms,  which provide unique spatio-temporal resolution through microchannel plate detection, were  particular fruitful in developing atom optics experiments.  Using both bosonic and fermionic helium species~\cite{Schellekens2005,Jeltes2007} these experiments observed the Hanbury Brown and Twiss (HBT) effect~\cite{Glauber1963b,Scully,Meystre,Loudon}.

More recently it was  demonstrated that metastable helium can be used also to create a direct source of atom pairs for the HOM effect~\cite{Lopes2015}.  The experiment started with a Bose Einstein condensate of helium atoms in the metastable $1s2s ^3S_1$ internal state.  A moving optical lattice created twin atom pairs in analogy to spontaneous four-wave mixing of optical fields [Fig.~\ref{exp}(c)].  Using another optical lattice, the modes for each of the twin beams were coherently superposed on an effective beam-splitter via Bragg diffraction [Fig.~\ref{fig:beamsplitter}(c)]. The timing of this effective beam-splitter process tuned the temporal overlap, and, hence, the atom distinguishability in analogy to the physical configuration of the original photonic HOM experiment. In those experiments, as  in the optical tweezer and lattice-based experiments, the presence of the HOM effect required  the realization of an effective beam-splitter and the preparation of indistinguishable bosonic atoms.  Recent experiments have also expanded to two particles and four modes~\cite{Dussarrat2017}.


\section{Explanation of the Hong-Ou-Mandel effect and many-particle interference}  \label{sec:HOMEffect}

One of the points we would like to make clear in our discussion of the HOM effect is the relation both to the statistical bosonic enhancement of multiply occupied single-particle states \emph{and} the coherent superposition of paths.  In this light, we will introduce the HOM effect by first discussing single-particle interference, then two-particle interference of the HOM effect, and then many-particle interference.

The superposition principle in quantum mechanics is most immediately illustrated using  a single particle that is prepared in a superposition of two distinct spatial states, supported by different modes - for example, in a double-slit experiment or in a Mach-Zehnder interferometer.
Although the superposition principle is typically discussed in the context of single-particle states,
it applies equally well to states of two or more particles:
if
$\ket{\Psi_i}\sim a_{i_1}^\dagger\hdots a_{i_n}^\dagger\ket{0}$ and
$\ket{\Psi_j}\sim a_{j_1}^\dagger\hdots a_{j_n}^\dagger\ket{0}$
are two $n$-particle states,
then so is any coherent superposition of these two states.
Just like coherent superpositions of single particle states give rise to interference that manifests itself in single particle observables,
coherent superpositions of $n$-particle states give rise to interference phenomena in $n$-particle space.
The most immediate paradigm for an observable consequence of many-particle interference is the HOM effect~\cite{HOM}, in which correlations in particle-like events are governed by wave-like interference in the high-dimensional many-particle space. Many-particle interference comes hand-in-hand with the particular particle statistics; typically, bosons interfere in a different way than fermions do.

\subsection{One particle:  Single-particle interference}

To develop some intuition, let us first establish some basic ideas and concepts of single particle interference in a Mach-Zehnder-interferometer and subsequently move to two-particle interference.
Initially, a particle in the optical mode $a_1$ is prepared in the state
\eq
\hat a^\dagger_1 \ket{0} , \label{HOM_Ini}
\en
where $\ket{0}$ denotes the vacuum and $\hat a_1^\dagger$ is a particle creation operator.
The particle impinges on a beam-splitter, which induces the time-evolution
\eq
\hat a^\dagger_1 \rightarrow  \hat U \hat a^\dagger_1 \hat U^{-1} = \sqrt{T} \hat b^\dagger_{2} + i \sqrt{R} \hat b^\dagger_1 ,  \label{timeevo}
\en
where $R$ and $T$ are the reflection and transmission coefficients of the beam-splitter connecting input modes $\hat a_1, \hat a_2$ with output modes $\hat b_1, \hat b_2$. The phase shift acquired upon reflection is chosen to be $\pi$.

Due to the unitarity of the time-evolution $\hat U$ induced by the optical element,
reflectivity $R$ and transmittivity $T$ satisfy $R+T=1$.
The beam-splitter thus generates a coherent superposition of a particle in the upper and a particle in the lower arm,
\eq
\left( \sqrt{T} \hat b^\dagger_{2} + i \sqrt{R} \hat b^\dagger_1  \right) \ket{0} ,
\en
where we exploited the invariance of the vacuum under particle-number-preserving time-evolutions, $\hat U \ket{0} \propto \ket{0}$.

Subsequently, the particle in the lower arm passes a phase shifter, such that it acquires a phase $\phi$ relative to the upper arm,
\eq
\hat b^\dagger_2 \rightarrow e^{i \phi} \hat b^\dagger_2 ,
\en
and the state becomes
\eq
\left( e^{i \phi}  \sqrt{T} \hat b^\dagger_{2} + i \sqrt{R} \hat b^\dagger_1  \right) \ket{0}\
\en

At a second beam-splitter, the two arms are joined again, and the time-evolution
\begin{eqnarray}
\hat b^\dagger_1 & \rightarrow & \sqrt{T} \hat c^\dagger_{2} +i  \sqrt{R} \hat c^\dagger_1\ ,\\
\hat b^\dagger_2 & \rightarrow & \sqrt{T} \hat c^\dagger_{1} + i  \sqrt{R} \hat c^\dagger_2\ ,
\end{eqnarray}
(similarly to Eq.~\ref{timeevo}) eventually leads to the final state
\eq
\left(  \left( - R + T e^{i \phi} \right) \hat c_1^\dagger +  i \left( \sqrt{RT} +\sqrt{RT} e^{i \phi}  \right) \hat c_2^\dagger \right) \ket{0}\,
\en
from which we can read off the probabilities $P(1,0)$ and $P(0,1)$ to find the particle in the upper (first) and in the lower (second) arm respectively.

These probabilities
\eq
P(1,0) & = &  R^2 +T^2 - 2 RT \cos (\phi) \\
P(0,1) & = &  4 R T \cos^2( \phi/2)
\en
depend on  the relative phase $\phi$, acquired between the two arms, since the amplitude for the pertinent event is the sum of the two complex path amplitudes.

\begin{figure}
\includegraphics[width=\linewidth,angle=0]{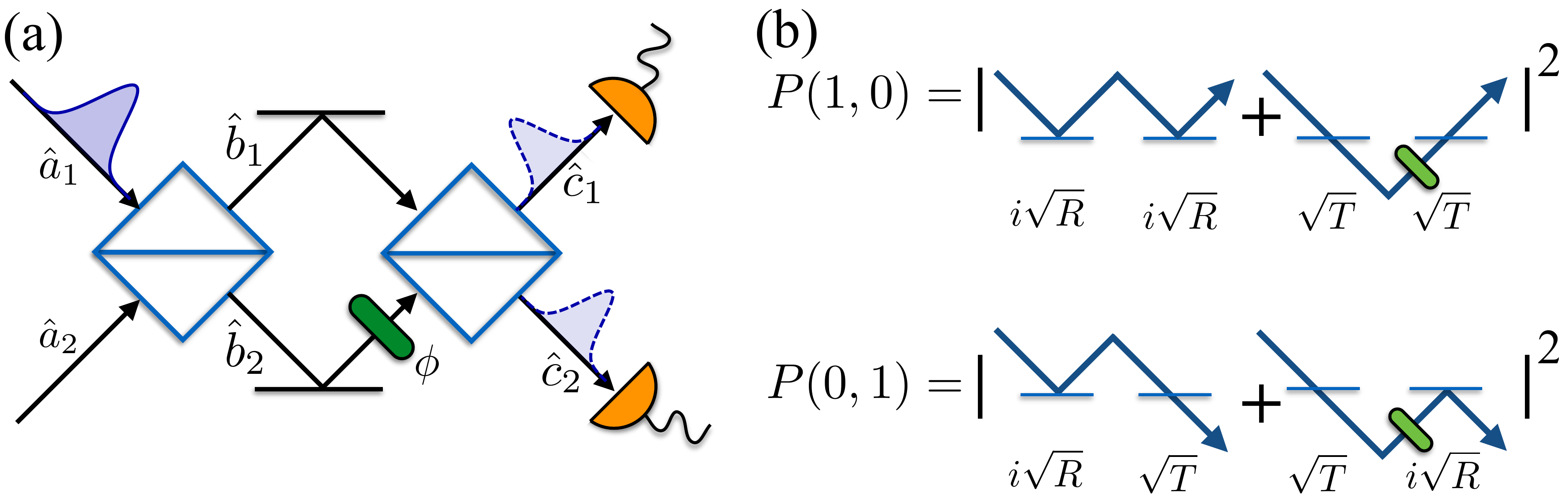}
\caption{Single-particle interference in the Mach-Zehnder interferometer. (a) Setup of two beam-splitters and mode labels. A relative phase $\phi$ is acquired in the lower arm, leading to interference fringes in the probability to find the particle in the upper output arm $\hat c_1$. (b) The probabilities for the events can be understood as arising from the coherent sum of single-particle path amplitudes. They reflect the acquired phase $\phi$.
\label{fig:doublebeamsplitter}}
\end{figure}

That is, two physically distinguishable, alternative pathways (the particle travels through the lower mode, the particle travels through the upper mode) are coherently super-imposed, leading to interference fringes as a function of $\phi$. As a result, the probability to observe a granular particle-like event is governed by the wave-like interference of the two pathways.

\subsection{Two-particle interference and the HOM effect}
\label{sec:two_particle_observables}

The HOM effect can be understood as an interference effect, but is inherently not a single particle effect as the interfering paths describe more than one particle.
Indeed, the HOM effect can not be understood in terms of single-particle interference.

Instead of a single particle, we now have to consider two identical particles in the two input modes of a beam-splitter:
\eq
\ket{\Psi_{\text{HOM,in}}} =
\hat a^\dagger_1  \hat a^\dagger_2 \ket{0}  \label{HOMinistate}
\en
and  detectors placed at the output modes of the beam-splitter as depicted in Fig.~\ref{HOMSketch}(a).

Because the particles do not interact, they propagate independently, and the beamsplitter is characterized by single particle dynamics similar to Eq.~\ref{timeevo}.
Since the HOM setup leaves no room for an additional phase shifter as in the Mach Zehnder setup,
we will leave the phase that a particle acquires upon reflection a free parameter.
The action of the beamsplitter thus reads
\begin{eqnarray}
\hat a^\dagger_1 & \rightarrow & \sqrt{T} \hat b^\dagger_{2} +i e^{i\varphi} \sqrt{R} \hat b^\dagger_1\ ,\\
\hat a^\dagger_2 & \rightarrow & \sqrt{T} \hat b^\dagger_{1} + i e^{-i\varphi} \sqrt{R} \hat b^\dagger_2\ .
\label{eq:beamsplitterphi}
\end{eqnarray}
The relative phase of $\varphi$ can be chosen at will, but unitarity prevents the two relative phases in the relations for $\hat a^\dagger_1$ and $\hat a^\dagger_2$ to be chosen independently.

With this, the final state of the particles after passing the beamsplitter reads
\eq
\left(  i\sqrt{R T} \left[ e^{i\varphi} \left(\hat b_1^\dagger\right)^2 + e^{-i\varphi} \left(\hat b_2^\dagger\right)^2 \right]  + T \hat b_2^\dagger \hat b_1^\dagger - R \hat b_1^\dagger \hat b_2^\dagger  \right) \ket{0}. \nonumber
\en

Up to here, the discussion applies to both fermions and bosons,
but in identifying probabilities $P(2,0)$, $P(0,2)$ to find both particles in the upper or lower arm,
or the probability $P(1,1)$ to find one particle in each arm,
one needs to distinguish between the two types of particles.

In the case of bosons,
the final state reads
\eq
i\left( \sqrt{R T} \left[ e^{i\varphi}\left(\hat b_1^\dagger\right)^2 + e^{-i\varphi} \left(\hat b_2^\dagger\right)^2 \right]  + (T-R) \hat b_1^\dagger \hat b_2^\dagger\right)\ket{0}, \nonumber
\en
due to the relation $b_1^\dagger \hat b_2^\dagger=b_2^\dagger \hat b_1^\dagger$. The state remains normalized to unity thanks to the properties of bosonic creation operations, which will be discussed in more detail below.  

Focusing  on the final state with one particle per output mode, $\hat b_1^\dagger \hat b_2^\dagger \ket{0}$, one can -- just like in the Mach-Zehnder interferometer -- identify two physically distinct paths that connect the initial and the final states:
Either both particles are reflected and feed the component $- R \hat b_1^\dagger \hat b_2^\dagger$, or both particles are transmitted and contribute to $T \hat b_2^\dagger \hat b_1^\dagger$. Interestingly, the phase relation between the quantum amplitudes is fixed to $-1=e^{i \pi}$, independently of the phase shift that a single particle acquires upon reflection.
Unitarity requires that the phase shift between a reflected and a transmitted particle is fixed. Moreover, any phase applied in any input mode only contributes as a global phase.  Hence,
the setup leaves no room to acquire a variable relative phase.

In the case of fermions, all contributions of doubly occupied states vanish since
$\left(\hat b_1^\dagger\right)^2=\left(\hat b_2^\dagger\right)^2=0$,
and the final state reads
\eq
-(T+R) \hat b_1^\dagger \hat b_2^\dagger \ket{0}=-\hat b_1^\dagger \hat b_2^\dagger \ket{0}
\en
because of the relation $b_1^\dagger \hat b_2^\dagger=-b_2^\dagger \hat b_1^\dagger$.
Independent of transitivity and reflectivity of the beamsplitter, one will thus always find one particle per mode.

The interference between the two discussed pathways is thus constructive in the case of fermions and destructive in the case of bosons.
Perfect destructive interference for bosons is achieved for $R=T=1/2$,
which is the case that gives rise to the HOM effect \cite{HOM}.

It should also be stressed that these interference effects are manifest in the probability to find two particles in the same mode or to find one particle in each mode.
These probabilities are reflected by two-particle observables, i.e. observables made of two creation and two annihilation operators, namely $\hat{b}_1^\dagger \hat{b}_2^\dagger \hat{b}_1 \hat{b}_2 $, $\left( \hat{b}_1^\dagger \right)^2 \left( \hat{b}_1 \right)^2 $,  $\left( \hat{b}_2^\dagger \right)^2 \left( \hat{b}_2 \right)^2 $. Proper single-particle observables (containing only one creation and one annihilation operator) like the expectation value for the number of particles in one mode, i.e.~$\hat{b}_1^\dagger \hat{b}_1$ and $\hat{b}_2^\dagger \hat{b}_2$, yield expectation values of unity for both bosons and fermions, independently of the properties of the beamsplitter, and they show no signature of interference.

\subsection{Many-particle interference beyond HOM:  More particles or more modes}
\label{sec:manyparticlephotons}

In its simplicity, the two-particle, two-mode HOM effect is a paradigm for the quantum behavior of a few indistinguishable particles as it reduces the essence of quantum statistics to a microscopic setting. Ultimately, this microscopic behavior is driven by the same cause as the macroscopic behavior of liquids, solids and gases. By increasing the number of particles and modes involved in the experiment~\cite{Ou:2007ly}, however, one encounters various surprising aspects, due to the interplay of many-body complexity and many-body coherence~\cite{Brunner2017,Dufour2017}.

\begin{figure}
 \includegraphics[width=\linewidth,angle=0]{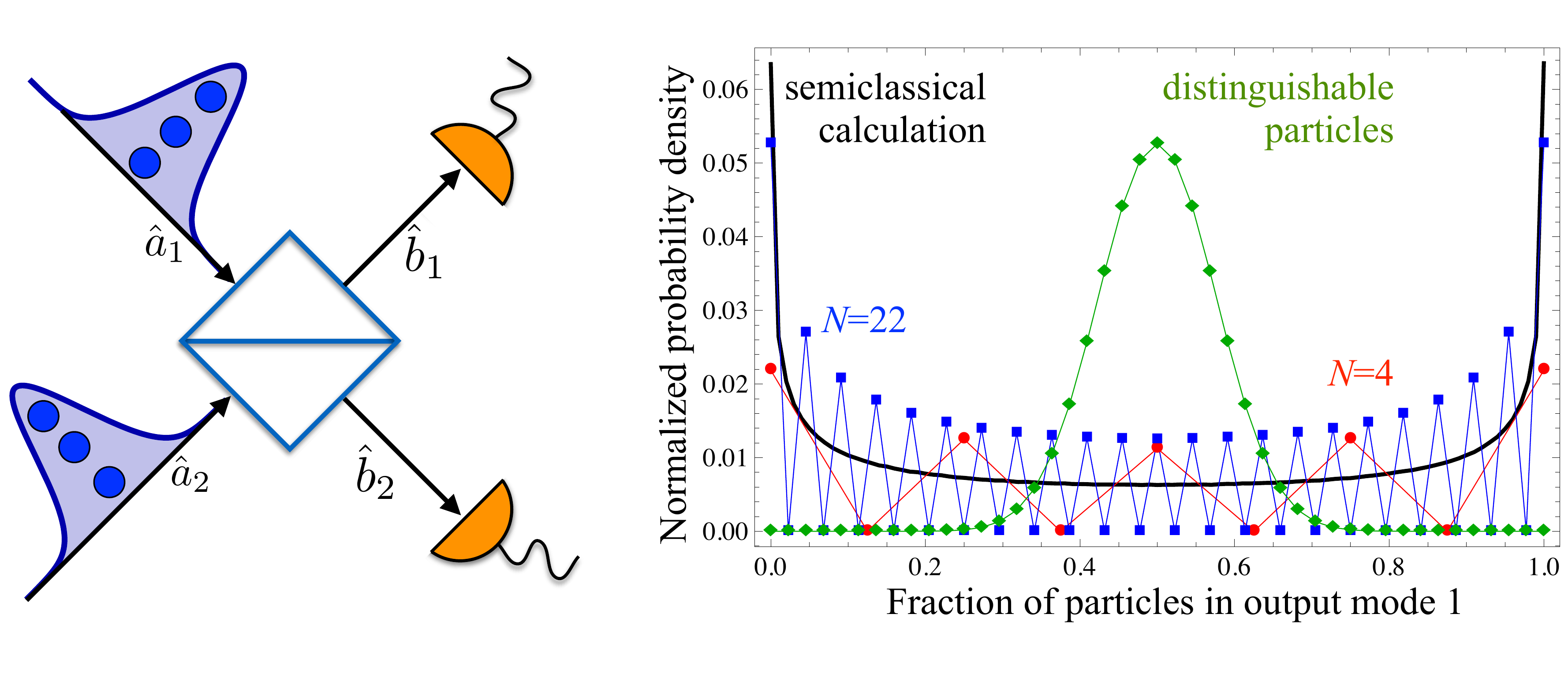}
 \caption{Two-mode many-particle scattering example.  Experimental setup on the left and calculation on the right. For many distinguishable particles that impinge on a beamsplitter, we expect to find a binomial distribution (green diamonds). For $N=4$ (red circles) and $N=22$ (blue squares) bosons that fall onto both input modes of a beamsplitter, the resulting interference pattern exhibits granular even-odd-like patterns, as well as a coarse-grained overall structure, which is reproduced well by a semi-classical approach (black solid line). The abscissa shows the fraction of particles encountered in one output mode, the ordinate is the respective probability, which was normalized such that it matches the continuous probability distribution of the semi-classical approach. Figure taken from Ref.~\cite{TichyThesis}.}
 \label{fig:twomodes}
 \end{figure}

The behavior of many identical bosons falling onto the two input modes of a balanced beamsplitter exhibits, both granular interference on the level of individual particles, as well as coarse-grained structures that can be described via a semiclassical approach \cite{Mullin:2006fk,Laloe:2010uq}, in which the number of particles is represented by a macroscopic wavefunction without well-defined phase. As a prominent example for a granular interference effect, consider the same number of particles  impinging onto a beam-splitter, i.e.~an initial state of the form
\eq
\left( \hat a_1^\dagger \right)^{N} \left( \hat a_2^\dagger \right)^{N} \ket{0}.
\en
The time-evolution is directly inherited as in Eq.~\ref{eq:beamsplitterphi}. As a direct generalization of the suppression of the transition $(1,1) \rightarrow (1,1)$, the probability to find an odd number of particles in either output mode vanishes \cite{Campos:1989fk,nockdipm,Laloe:2010uq}, which can be shown for the general case by symmetry considerations.  In one experiment studying multi-particle interference, the suppression of the transition $(2,2) \rightarrow (3,1)$ was accompanied with a narrowing of the temporal width of the interference due to multi-particle effects~\cite{younsikraNatComm}. On top of this fine-grained interference pattern, a coarse-grained structure appears: When many particles impinge on both modes of the beamsplitter, the probability to find a balanced number in the output modes is suppressed, and unbalanced arrangements are enhanced \cite{Laloe:2010uq}. The interplay of fine-grained and coarse-grained structures can be seen in Fig.~\ref{fig:twomodes}.

When increasing the number of modes as well as the number of particles, the number of possibilities to arrange the particles in the modes explodes combinatorially. On the other hand, the freedom in choosing a unitary scattering matrix $A$ that governs the many-mode time-evolution
\eq
\hat b_n \rightarrow \sum_{m=1} A^n_m \hat a_m,
\en
in analogy to Eq.~\ref{eq:beamsplitterphi} allows the formulation of very different experiments \cite{Reck:1994zp}. While general statistical tendencies -- bosons prefer bunched arrangements over spread-out configurations while the Pauli principle for fermions allows only such spread-out configurations \cite{Tichy:2012NJP} -- prevail in all configurations, the very computation of individual transition probabilities becomes prohibitively difficult for bosons. Indeed, the transition probability becomes the \emph{permanent} of the respective scattering matrix \cite{Scheel:2004hc}, for which the computational complexity is well-established. As a consequence, even the mere simulation of many bosons that scatter off a random setup is a computationally hard problem, dubbed boson-sampling \cite{Aaronson:2010fk}, which has attracted wide interest, fueled also by its experimental feasibility for few photons \cite{Spring15022013,Tillmann:2012ys,Crespi:2012vn,Broome15022013}.

Whereas an unstructured random scattering setup does not allow systematic physical statements on the behavior of the particles in the scattering process, structured setups with symmetries allow the formulation of strong rules for the suppression of output events: If a symmetry is present in the input arrangement of bosons and the setup is symmetry-preserving, the output arrangements need to fulfill the symmetry as well. This principle was applied to the Fourier-matrix \cite{Lim:2005qt,Tichy:2010ZT,Tichy:2012NJP,Tichy:2013lq}, as well as to scattering matrices with other kinds of symmetries  \cite{PhysRevA.91.013811,Dittel2016}. The experimental implementations with up to five photons \cite{Crespi2016,Spagnolo:2013fk} may also provide one way to verify the functionality of an alleged boson-sampling computer \cite{Spagnolo:2013eu}.

\subsection{Bosonic bunching and fermionic anti-bunching}

We now turn to understanding the two-particle and many-particle interference in a number of contexts.  Returning to Fig.~\ref{HOMSketch}(b), note that independent of the particle species, a given final state with both particles in one mode (either the first or the second one) is fed by exactly one many-particle path.
One particle must be reflected, the other one is necessarily transmitted,
and no interference of different two-particle paths takes place.
Nevertheless, the probability to find both particles in one output mode does not coincide with $RT$ as one might have expected for an event in which one particle is transmitted and one is reflected. This result would hold true for distinguishable particles, but
the probability for bosons is enhanced by a bunching factor 2,
whereas coincidence events are completely suppressed for fermions.
The origin of this behavior is the state-space-structure of bosons and fermions. The Pauli principle for  fermions  strictly prohibits the multiple occupation of any single-particle state, in any basis,
\eq N >1 \Rightarrow  (\hat b^\dagger)^N \ket{0} = 0 ,
\en
while the bunching tendency of bosons is reflected by the over-normalization of multiply populated states,
\eq
\left( \hat b_1^\dagger \right)^2 \ket{0} = \sqrt{2} \ket{2} .
 \en
Probing the occupation of the final Fock-state $\ket{2,0}$ or $\ket{0,2}$, allows us to witness the fermionic suppresion and bosonic preponderance of double-occupied states -- but not the coherent superposition of several many-particle paths.

The suppression for fermions is by no means specific to the two-particle HOM setup, and applies to many-particle interference. Indeed, the probability to find several fermions in the same single-particle state is fully suppressed  in every thinkable setup -- the Pauli principle is universal \cite{Javorsek:2000bh,Collaboration:2010ly}.
Similarly, also the bosonic enhancement of a bunching event in which all particles end in the very same mode fulfills \cite{TichyThesis,PhysRevLett.111.130503,TichyMultiDimPerm}
 \eq
 P_B = P_D \frac{ n! }{\prod_{j=1}^l r_j! },
 \en
 where $P_B$ is the probability for $n$ bosons prepared with occupations $r_1, \dots, r_l$ to end in the same output mode, and $P_D$ is the analogous probability for distinguishable particles. The enhancement factor $n! / \prod_j r_j!$ is independent of any phases acquired in the system and reflects merely the bosonic state-space-structure. 
 

\subsection{Interference versus statistics}

Often, Dirac's famous dictum \cite{Dirac:1930vn}, ``Each photon then interferes only with itself. Interference between different photons never occurs.''~is said to be overcome in view of the HOM effect. However, it remains fully valid  in the present context: The first particle does not interfere with the second -- the two particles propagate independently and reach each output mode with a definite probability, independently of the presence of the other one.
Nevertheless, coherent superpositions of different two-particle states
results in interference,
but this interference is only apparent in two-particle observables as discussed in the specific case of the HOM effect above.
When probing any single-particle observable, e.g.~the average particle number in each output mode,
there is no difference between bosons, fermions or distinguishable particles

We have treated the two types of events -- doubly occupied bunching events (2,0) and coincident events (1,1) -- separately. Indeed, we witness two rather independent phenomena in one setup: The statistical bosonic enhancement or fermionic suppression of multiply occupied single-particle states \emph{and} the coherent superposition of many-particle paths. Of course, due to the conservation of probability, an enhancement of (1,1) needs to be accompanied by a suppression of (2,0), and vice versa, and it may seem suspicious to view bosonic/fermionic statistical behavior and many-particle interference as two separate physical phenomena. Admittedly, these phenomena have a common origin: the indistinguishability of particles of the same species. The dependence of the (2,0)-event on the (1,1)-probability is, however, not by any means universal, it is unavoidable in the basic two-particle HOM effect due to the small size of the setup: When increasing the number of modes \cite{PhysRevLett.108.010502,Gard:13} and/or the number of particles \cite{Laloe:2010uq}, statistical effects (bosonic bunching and fermionic suppression of multiply occupied events) and many-particle interference (governing coincident events) become quite independent phenomena. In particular, bosons and fermions can exhibit very similar interference patterns in single (or few) particle observables.
Many-particle interference on the other hand is much more sensitive to phase variations
and a meticulously prepared initial state than statistical bosonic or fermionic effects \cite{Tichy:2012NJP}.

\subsection{Loss of coherence and partial distinguishability}

Interference effects can generally be easily be washed out by noise or decoherence.  It is illustrative to understand the effect of loss of coherence at the beamsplitter on the HOM effect.  To provide an illustrative calculation, we consider here mainly the case of two-particle interference.  However, as just mentioned many-particle interference will exhibit different sensitivities.

Further, because the presently discussed interference effects are based on particle indistinguishability,
they are also easily affected by imperfect state preparations that make different particles distinguishable by some physical property. 

\subsubsection{Loss of coherence in the beamsplitter}
\label{sec:mixedstates}
The beamsplitter with the relation defined in Eq.~\ref{timeevo} induces a perfect phase relation between the two outgoing states; that is, the splitting is perfectly coherent.

In order to discuss reduction of phase coherence, it is instructive to consider a beamsplitter following Eq.~\ref{eq:beamsplitterphi}
that describes a coherent operation for any value of $\varphi$.
Ideally $\varphi$ is a constant, and in fact to utilize the HOM effect as a quantum resource $\varphi$ must be controlled, but in real experiments $\varphi$ may fluctuate.
In this section, we will briefly discuss the impact of such fluctuations on the HOM effect and entanglement properties of the final state. For further discussion of the mode entanglement associated with the HOM effect see Sec.~\ref{sec:entanglement_identical}, where we will assume a non-fluctuating phase can be verified.

The state resulting from the beamsplitter Eq.~\ref{eq:beamsplitterphi}
with a given phase $\varphi$ in an HOM experiment reads
%
\eq
(T-R)\ket{1,1}+i\sqrt{2RT}(e^{i\varphi}\ket{2,0}+e^{-i\varphi}\ket{0,2}).
\label{eq:pureRT}
\en
with $\ket{1,1}=\hat b^\dagger_1\hat b^\dagger_2\ket{0}$, $\ket{2,0}=\hat b^\dagger_1\hat b^\dagger_1\ket{0}$ and
$\ket{0,2}=\hat b^\dagger_2 \hat b^\dagger_2\ket{0}$.
For a balanced beamsplitter with $R=T=1/2$ one will thus always observe perfect bunching independently of the value of $\varphi$.

In the case of a shot-to-shot-fluctuating phase described in terms of a distribution $\mu(\varphi)$, the resulting mixed state (i.e.~ density matrix) averaged over the phase fluctuation reads
\eq
\varrho_\alpha=\left[
\begin{array}{ccc}
2RT & i\sqrt{2RT}Q\alpha & 2RT\alpha^2 \\
-i\sqrt{2RT}Q\alpha^\ast & Q^2 & -i\sqrt{2RT}Q\alpha \\
2RT(\alpha^\ast)^2 & i\sqrt{2RT}Q\alpha^\ast & 2RT
\end{array}
\right]
\label{eq:rhoalpha}
\en
in the basis $\ket{2,0}$, $\ket{1,1}$, $\ket{0,2}$
with
\eq
\alpha=\int d\mu(\varphi)e^{-i\varphi}\ ,
\en
and the short hand notation $Q=T-R$.
Because the probabilities for the observation of mode occupations is independent of the value of $\varphi$, they are also not affected by phase fluctuations.
That is, a perfect HOM dip can be observed even for maximal phase fluctuations for which $\alpha=0$. These phase fluctuations however have a strong impact on the entanglement properties of the final state~\cite{TichyJOP,natphys:345}.  Similarly, many-particle, many-mode interference will see detrimental effects from phase fluctuations.

\subsubsection{Partial indistinguishability}

If both particles impinging on the beamsplitter overlap perfectly, 
one observes the HOM effect discussed thus far,
but if both particles propagate through the beamsplitter one after the other, no such effect exists.  Similarly for atoms distinguished in another degree of freedom such as motion or spin, interference will not occur.
To discuss the range of cases in between, we can consider a situation in which the spatial wave functions of both particles do not overlap perfectly.
To this end, we can assume the particle in mode $1$ to be prepared in the state $\hat a_1^\dagger\ket{0}$,
and the particle in mode $2$ to be prepared in the state $(\cos\theta\ \hat a_2^\dagger+\sin\theta\ \tilde a_2^\dagger)\ket{0}$,
where $\tilde a_2^\dagger\ket{0}$ is a state that yields no overlap of the two particles in the beamsplitter at all.
Varying $\theta$ from $0$ to $\pi/2$ thus permits to continuously scan from perfectly indistinguishable particles to completely distinguishable particles.

It is instructive to first consider the case in which the two particles do not overlap at all, that is, the classical combinatorial case.
From the relation
\eq
\tilde a_2^\dagger \hat a_1^\dagger\to
T b_2^\dagger\tilde b_1^\dagger-R\tilde b_2^\dagger \hat b_1^\dagger+i\sqrt{RT}(e^{i\varphi}\tilde b_1^\dagger \hat b_1^\dagger+e^{-i\varphi}\tilde b_2^\dagger \hat b_2^\dagger)\nonumber
\en
one directly reads off the probability $RT$ to find both particles in the first or second output mode respectively,
and the probability $T^2+R^2$ to find one particle in each output mode.

In the case of a general input state
$(\cos\theta\ a_2^\dagger+\sin\theta\ \tilde a_2^\dagger)\ket{0}$
the corresponding probabilities for finite overlap read for bosons
\eq
P_B(2,0)=P_B(0,2)&=&
RT(1+\cos^2\theta)\\
P_B(1,1)&=&
(R-T)^2+2RT\sin^2\theta
\en
 and for fermions
\eq
P_F(2,0)=P_F(0,2)&=&\sin^2\theta\ RT\\
P_F(1,1)&=&\cos^2\theta+\sin^2\theta\ (R^2+T^2)\\
&=&R + T-2\sin^2\theta\ RT.
\en
Hence, the overlap of the wavefunctions can be read off in the event probabilities. This makes the HOM effect a direct quantitative probe for indistinguishability.


\section{Entanglement and the Hong-Ou-Mandel effect}

So far, we have examined how interference emerges when indistinguishable particles are superposed among two or many modes, emphasizing the role of multi-particle path interference. We showed that a multiparticle observable cannot be inferred from the expectation of single-particle interference, but rather depends on the quantum statistics of the particles (Sec.~\ref{sec:two_particle_observables}). When multiparticle observables display correlations that cannot be predicted by single-particle observables, often quantum entanglement plays an important role. However, one must be careful when considering entanglement of identical particles because in this circumstance the typical entanglement criterion of non-separability can be misleading.

In this section, we will explore the entanglement present in the context of the HOM effect and its generalizations. The HOM effect involves multiple particles, but also multiple modes, or paths.  We will introduce concepts we refer to as mode entanglement and particle entanglement~\cite{TichyJOP}, which illustrate the different ways non-separability are manifest in relevant quantum states.  In particular, we reserve the term particle entanglement for a situation in which, due to entanglement, a physical reality cannot be assigned to a \textit{particle} individually, and a salient point of our discussion is that no particle entanglement is present in any single or multi-particle HOM effect with spinless non-interacting particles.  We will first generally discuss entanglement and related concepts, then, we use these concepts to define entanglement for identical particles in a way that assures the presence of correlated fluctuations associated with the essence of entanglement.   Along the way we compare and contrast to prototypical Bell states in the particle's spin or polarization degree of freedom.  Bell state creation in atomic physics is mainly associated with entangling interactions, but, as we will describe, generalizations of the HOM effect also enable Bell state creation through measurement, effectively transferring entanglement between external (mode) and internal degrees of freedom.

\subsection{Entangled states}
\label{sec:entanglementpurestates}

As stated, our goal for this section is to discuss the varying roles of entanglement in the HOM effect, and we do so by first addressing entanglement generally. Entanglement gives rise to correlations of measurement results that can not be accounted for in terms of classical probabilities. A prototypical example of an entangled state is a Bell state.
It consists of two distinguishable particles, which we will assume have spatial positions $\scL$ and $\scR$, with internal states $\ket{\uparrow}$ and $\ket{\downarrow}$ that could for example correspond to the alignment of an internal degree of freedom along a quantizing direction. In this basis, the Bell state $\ket{\Phi_+}$ can be written,
\begin{equation}
\ket{\Phi_+} = \frac{1}{\sqrt{2}}\left (\ket{\uparrow}_\scL\ket{\uparrow}_\scR+\ket{\downarrow}_\scL\ket{\downarrow}_\scR\right ).
\label{eq:bellstate}
\end{equation}
Local measurements in the basis $\{\ket{\uparrow},\ket{\downarrow}\}$ have maximally undetermined outcomes, i.e.~the probabilities to project on $\ket{\uparrow}$ and $\ket{\downarrow}$ are both $50\%$.
Despite the maximal fluctuations in single-particle observables, once the state of one of the particles has been measured, the outcome of a subsequent measurement on the second particle is determined. One will always observe that the particles have the same spin state, and hence two-point observables (i.e.~correlations) are not fluctuating. Crucially, this is not limited to measurements in the bases $\{\ket{\uparrow}_\scL,\ket{\downarrow}_\scL\}$ and $\{\ket{\uparrow}_\scR,\ket{\downarrow}_\scR\}$. For any basis in $\scL$ (i.e.~any unitary rotation of the basis vectors), there is a corresponding basis in $\scR$ such that measurement results are perfectly correlated, while results of each measurement alone remain completely undetermined. While classically such correlations could be produced in a particular basis, it is only through quantum effects that correlations can persist in multiple bases.

Entangled states are often formally expressed as a state that is not separable. A separable pure state is one which can be written as,
\begin{equation}
\label{eq:sep}
\ket{\psi} = \ket{\psi}_\scL\otimes\ket{\psi}_\scR
\end{equation}
of subsystem states $\ket{\psi}_\scL$ and $\ket{\psi}_\scR$. Any pure state that cannot be written in this way, such as the state Eq.~\ref{eq:bellstate} with respect the spin-1/2 subsystems at $\scL$ and $\scR$, is usually considered entangled. As we will discuss in what follows, identical particles can complicate this somewhat simple diagnostic formula.

\subsection{Symmetrization and separability}
\label{HOM:Id_Ent}

Sec.~\ref{sec:entanglementpurestates} treats the particles as distinguishable. The existence of two degrees of freedom (e.g. spatial position and spin) allows reference to the particle at position $\scL$ and the one at $\scR$, each with its own spin degree of freedom. As such, the quantum statistics of the particles do not have physical consequences. By contrast, in the HOM effect the bosonic particles have a single degree of freedom (spatial position, i.e.~mode), and the quantum statistics are essential.

By writing out the quantum states involved in the HOM effect, the complications that arise from identical particles are apparent. Adopting a first quantized notation, we will refer to the particles in the two-particle state as $i$ and $ii$. In general, if we have a bosonic state with a particle in state $\ket{\alpha}$ and the other $\ket{\beta}$ (with $\langle \alpha \ket{\beta} = 0$), then the symmetrized state is written,
\begin{equation}
\ket{\psi_{\rm sym}} = \frac{1}{\sqrt{2}}\left (\ket{\alpha}_{i}\ket{\beta}_{ii}+\ket{\beta}_{i}\ket{\alpha}_{ii} \right),
\label{eq:symmState}
\end{equation}
which is manifestly symmetric under exchange of the particle labels $i$ and $ii$. Although this describes the simple scenario of two particles in orthogonal quantum states, it exhibits non-separability with respect to the particle labels $i$ and $ii$. This raises the question of whether the non-separability constitutes entanglement. The goal of the following sections is to bring further clarity to this point, namely, how the non-separability intrinsic to all multi-particle state of identical particles should be squared with entanglement.

\subsection{Entanglement of identical particles}
\label{sec:entanglement_identical}

To analyze the presence of entanglement in Eq.~\ref{eq:symmState}, we first appeal to the more basic notion of entanglement developed in Sec.~\ref{sec:entanglementpurestates}. The state $\ket{\Phi_+}$ exhibits correlated fluctuations in the spin basis associated with spatially-separated particles. The state $\ket{\psi_{\rm sym}}$, on the other hand, does not exhibit any fluctuations: an observer always measures a particle in state $\ket{\alpha}$ and another particle in state $\ket{\beta}$. In the state $\ket{\Phi_+}$, the subscripts representing position and the ket arguments representing spin correspond to the eigenstates of observable operators. By contrast, the particle label subscripts $i$ and $ii$ in $\ket{\psi_{\rm sym}}$ are not physical as they do not correspond to the eigenstate of any observable. Hence, although the state $\ket{\psi_{\rm sym}}$ is non-separable, it lacks the physical properties associated with the state $\ket{\Phi_+}$. This suggests separability alone is an insufficient metric for identifying entanglement where identical particles are concerned.

\subsubsection{Complete set of properties (CSOP) analysis}

There is a framework that clarifies the separability conundrum, which is described in Refs.~\cite{TichyJOP, TichyThesis, ItalianPRA}; we closely follow the treatment of these references here. To motivate the idea behind the framework, one must relax the constraint of separability for non-entangled states, but retain the more fundamental quantum-correlated fluctuations that underlie entanglement.

The framework we summarize is rooted in the following idea.  When two particles are non-entangled, one can ascribe a ``complete set of properties", i.e.~physical reality, to each of the particles, irrespective of which particle it is that carries each of these sets.  For the state in Eq.~\ref{eq:symmState},  one of the particles is always in state $\ket{\alpha}$ and the other in $\ket{\beta}$. Even though formally the unphysical particle label associated with the states ($\ket{\alpha}$ or $\ket{\beta}$) fluctuates, there are no measurable fluctuations in any physical observable, and, as such, no entanglement between the particles.

Reference~\cite{ItalianPRA} provides a formalism for identifying whether a particle within a two-particle quantum state has a complete set of properties (CSOP). The formalism relies on the construction of the symmetric operator $\mathcal{E}_P$, with unity expectation value when one of the particles in a symmetrized two-particle state has a CSOP. To develop the formalism, consider the single-particle Hilbert space spanned by the orthonormal set $ \{x_j\}$. If a Bose-symmetrized state $\ket{\psi_{\rm sym}}$  consists of particles with a CSOP, then there is a projector in the single-particle Hilbert space $P = \ket{x_j}\bra{x_j}$, satisfying~\cite{ItalianPRA,TichyThesis},
\begin{equation}
\label{HOM:id_part_Proj}
1 = \bra{\psi_{\rm sym}}  \mathcal{E}_P \ket{\psi_{\rm sym}},
\end{equation}
with,
\begin{equation}
 \mathcal{E}_P = P\otimes (\mathbb{I}-P)+ (\mathbb{I}-P)\otimes P,
\label{HOM:id_part}
\end{equation}
where the Kronecker product $\otimes$ is taken with respect to the single particle Hilbert spaces associated with each particle. When these equations are satisfied, one of the particles has a CSOP associated to the state $\ket{x_j}$, and the correlated fluctuations intrinsic to entanglement cannot be present. The other particle is always in a state $\ket{x}$ (in general, a superposition of states in the orthonormal set excluding $\ket{x_j}$) orthogonal to $\ket{x_j}$, satisfying $1 = \bra{x} \mathbb{I}-P\ket{x} = \bra{x}\sum_{l\neq j} \ket{x_l}\bra{x_l} x \rangle$. However, because two bosons can share the same quantum-state, and yield $0 =\bra{\psi_{\rm bos}}  \mathcal{E}_P \ket{\psi_{\rm bos}}$, we must provide the additional prescription that a non-entangled state either satisfies Eq.~\ref{HOM:id_part_Proj} or describes two bosons occupying the same quantum state. If neither is true, the particles are entangled. The criteria can be summarized as follows: two bosons are not entangled when either they share the same quantum state, or are the symmetrized product of two orthogonal quantum states~\cite{TichyThesis,ItalianPRA, TichyJOP}. While the projection operator $ \mathcal{E}_P$ exists for the exemplary state $\ket{\psi_{\rm sym}}$ of Eq.~\ref{eq:symmState}, there is provably no such operator for $\ket{\Phi_+} $ and so it exhibits ``particle entanglement''. Lastly, we show in Sec.~\ref{ref:Schmidt} that the prescription here in terms of the symmetric operator $ \mathcal{E}_P$ can be equivalently cast in terms of the Schmidt rank, which also provides a useful tool for understanding the effect of interactions.

\subsubsection{Applying CSOP analysis to HOM}

We now apply the CSOP formalism to elucidate the entanglement in the input and output state in the HOM experiment. We will consider two kinds of entanglement -- that we refer to as mode and particle entanglement -- and whether each is manifest in the HOM effect. To do so, we will go between a first quantized and second quantized notation where appropriate to the discussion.

We start with first quantization in order to apply the treatment of Sec.~\ref{HOM:Id_Ent} to the HOM effect. We consider two spinless bosons that initially occupy distinct spatial modes. These modes could be a propagating free-space mode of a photon or atom, or the bound state of a harmonic potential. 
 We will adopt the latter scenario of a double well with bound states (summarized in Sec.~\ref{sec:double_well}) for the following discussion. We will call the bound state modes $\ket{\scL}$ and $\ket{\scR}$. The initial two-particle state is,
\begin{equation}
\ket{S} = \frac{1}{\sqrt{2}}\left (\ket{\scL}_i\ket{\scR}_{ii}+\ket{\scR}_{i}\ket{\scL}_{ii} \right ).
\end{equation}
We can now ask whether the particles are entangled according to the metric established in Section~\ref{HOM:Id_Ent}. We seek the projector $P$ that satisfies Eqs.~\ref{HOM:id_part_Proj} and~\ref{HOM:id_part}. Trivially, if we take $P = \ket{\scR}\bra{\scR}$, then $\bra{S}  \mathcal{E}_P \ket{S} = 1$. The input state $\ket{S}$ describes particles that are not entangled, because it lacks any measurable quantum fluctuations that might support correlations~\cite{ItalianPRA}.

We now consider the output state after HOM interference. In a double well, a tunnel-coupling between two bound states realizes a beam-splitter interaction, inducing the single-particle transformations of the form (irrespective of the particle label),
\begin{equation}
 \{ \ket{\scL}, \ket{\scR} \} \rightarrow \{\frac{1}{\sqrt{2}} \left (\ket{\scL}+i\ket{\scR} \right ), \frac{1}{\sqrt{2}} \left (\ket{\scR}+i\ket{\scL} \right ) \},
\end{equation}
which when implemented on $\ket{S}$ yields,
 \begin{align}
 \ket{S} &\rightarrow \frac{1}{\sqrt{2}} \left( \ket{S} - \ket{S} + i\ket{\scL}_i\ket{\scL}_{ii} +i\ket{\scR}_i\ket{\scR}_{ii} \right) \nonumber \\
 &\qquad {} =  \frac{i}{\sqrt{2}} \left ( \ket{\scL}_i\ket{\scL}_{ii} +\ket{\scR}_i\ket{\scR}_{ii} \right) \nonumber \\
 &\qquad {} \equiv \ket{+},
 \label{beamTrans}
 \end{align}
 where in the last line we have introduced a notation for the ouput state, $\ket{+}$.  We now analyze $\ket{+}$ for the presence of particle entanglement. Again, from the orthonormal space $\{ \ket{\scL},\ket{\scR} \}$ we seek the pure-state projector $P=\ket{p}\bra{p}$ that satisfies Eqs.~\ref{HOM:id_part_Proj} and~\ref{HOM:id_part}. Importantly, $\ket{p} = \frac{1}{\sqrt{2}} \left (\ket{\scL} + i\ket{\scR} \right )$ is sufficient, leading to projectors in $\mathcal{E}_p$ that correspond exactly to the beam-splitter transformed states. As in $\ket{\psi_{\rm sym}}$ in Eq.~\ref{eq:symmState},  the beamsplitter  results in $\ket{\alpha}= \frac{1}{\sqrt{2}} \left (\ket{\scL}+i\ket{\scR} \right )$ and  $\ket{\beta}=\frac{1}{\sqrt{2}} \left (\ket{\scR}+i\ket{\scL} \right )$. One of the particles is always in the spatial superposition given by $\ket{\alpha}$, and the other in the superposition given by $\ket{\beta}$, and the corresponding symmetrized two-particle state is $\ket{+}$. According to the prescription from Section~\ref{HOM:Id_Ent} the existence of $P$ implies that there is not particle entanglement in the state $\ket{+}$. Hence, neither the input or output state in the HOM effect requires particle entanglement.

Lastly, we stress that the usage of a double well in the above example does not restrict the generality of the argument. In particular, the double well has the feature that the input modes and output modes are the same (see Fig.~\ref{fig:beamsplitter}). However, had we redefined the output modes in the above to be distinct from the input modes, the final quantum state would still lack particle entanglement according to the CSOP prescription.

\subsubsection{Second quantization and mode entanglement}

While the first-quantized picture demonstrates an absence of particle entanglement in the state generated in the HOM experiment, a second-quantized treatment emphasizes the presence of mode entanglement~\cite{Ray2011}. As described in Sec.~\ref{sec:HOMEffect}, we are interested in the initial state of two particles occupying two distinct input modes, with associated creation operators $\hat a_1^\dagger$ and $\hat a_2^\dagger$. As in Eq.~\ref{HOMinistate}, the initial state of two modes each with a single excitation is $\ket{\Psi_{\text{HOM,in}}} = \hat a_1^\dagger \hat a_2^\dagger\ket{0} = \ket{1,1}$. This state is physically equivalent to the state $\ket{S}$ from the first quantization picture. On the other hand, the output state after a balanced beamsplitter is $\ket{\Psi_{\text{HOM,out}}} = \frac{1}{2}((\hat b_1^\dagger)^2 +(\hat b_2^\dagger)^2)\ket{0} = \frac{1}{\sqrt{2}}(\ket{2,0}+\ket{0,2})$, which is equivalent to the state $\ket{+}$. Note if a single photon were to impinge on one port of the beamsplitter, the output state $\frac{1}{\sqrt{2}}(\ket{1,0}+\ket{0,1})$ would be similarly mode entangled.

In this second-quantized notation, we now ask whether there is separability between the two physically observable mode degrees of freedom and the Fock state quantum number within each mode. The initial state $\ket{\Psi_{\text{HOM,in}}}$ is a product state with respect to these degrees of freedom, while the output state is not. Therefore, the initial state lacks mode entanglement, while the output state acquires it due to the application of the beamsplitter. We also stress that there is no logical inconsistency in applying separability in this context, because we are asking about separability between physically observable degrees of freedom, whereas the separability in question within a generic symmetrized state (Eq.~\ref{eq:symmState}) is related to the unphysical particle labels.



\subsection{Schmidt Rank analysis}
\label{ref:Schmidt}
As shown in Ref.~\cite{ItalianPRA}, the CSOP prescription can be linked to the Schmidt rank, which quantifies entanglement between the sub-systems of a quantum state. The Schmidt rank provides an intuitive partition between how mode and particle entangled states are generated. Quite sensibly, mode-entangled states can only result when there are interactions between the modes, and particle-entangled states can only result when there are interactions between the particles (or projective measurement).

The Schmidt rank treatment discussed here pertains strictly to bosons, but can be modified to the case of fermions. We consider the case of two bosons, 1 and 2, represented in a basis of orthonormal single-particle states $\{ \ket{x_j} \}$. We use a first quantized picture in order to emphasize the presence of symmetrization and the emergence of particle entangled states. In general, any Bose-symmetrized two-particle state can be written in first quantization as,
\begin{equation}
\ket{\psi_{\rm 2p}} = \sum_{q,j} v_{q,j} \ket{x_q}_i\ket{x_j}_{ii},
\label{orthogonal}
\end{equation}
where  $v_{q,j}$ is the amplitude associated with particle 1 being in $\ket{x_q}$ while particle 2 is in $\ket{x_j}$. This is the traditional representation, for example, for a pair of particles occupying two orthogonal quantum states. Because these are bosons, symmetrization implies $v_{q,j} = v_{j,q}$. The matrix $v$ (with elements $v_{q,j}$) can be diagonalized~\cite{ItalianPRA}, such that we can rewrite the two-particle wave function in the diagonal basis $\{\ket{\bar{x}_j}\}$,
\begin{equation}
\ket{\psi_{\rm 2p}} = \sum_{k} \bar{v}_{k} \ket{\bar{x}_k}_1\ket{\bar{x}_k}_2,
\label{diagonalized}
\end{equation}
where $\bar{v}_k$ is the amplitude associated with both particles occupying the same state $\ket{\bar{x}_k}$. To discriminate the presence of particle entanglement, we introduce the ``Schmidt rank'', which is the number of terms in the summation with non-zero $\bar{v}_{k}$. The particles are entangled when the Schmidt rank exceeds two~\cite{ItalianPRA, TichyJOP}.  When the Schmidt rank is two, the state is the symmetrized product of two quantum states (not necessarily orthogonal); when the Schmidt rank is one, the particles are in the same exact quantum state.

The Schmidt rank treatment can be straightforwardly applied to the HOM effect. The input state $\ket{S}$ has Schmidt rank of two, with $\{ \ket{\bar{x}_\pm} \} = \frac{1}{\sqrt{2}} (\ket{\scL} \pm \ket{\scR})$. This rank is preserved by the unitary transformation achieved by the beamsplitter to produce $\ket{+}$, whose first-quantized description is in the format of Eq.~\ref{diagonalized} (i.e.~already expressed in the diagonal basis). Both the input and output two-particle states are the symmetrized product of orthogonal single-particle quantum states. Hence, there is no particle entanglement in the HOM effect.  This is to be expected; if one starts with a pair of non-particle-entangled bosons in two orthogonal quantum states, then the action of a single-particle Hamiltonian (i.e.~no interactions) will leave the state in that form, because the unitary Schrodinger evolution preserves the orthogonality of the evolving states. In other words, the two-particle state remains a symmetrized product of two orthogonal quantum states, and the Schmidt rank of two is invariant with time.


\subsection{Creating particle entanglement through measurement or interactions}

When measurement or interactions are at play, it is possible to produce a two-particle quantum state whose Schmidt rank exceeds two, and hence, contains particle entanglement. For example, consider the state $\ket{\Phi_x}$ below, which is related to $\ket{\Phi_+} $ by basis rotation. This state is routinely produced with ions via their Coulomb interaction~\cite{Sackett2000} and single-particle spin rotations. In first quantization, for bosons this state is written,
\begin{equation}
\ket{\Phi_{x}} = \frac{1}{2}\left (\ket{\scL}_i\ket{\scR}_{ii}+\ket{\scR}_i\ket{\scL}_{ii} \right ) \left (\ket{\uparrow}_i\ket{\downarrow}_{ii}+\ket{\downarrow}_i\ket{\uparrow}_{ii} \right).
\end{equation}
In the basis $\{ \ket{x_j} \} = \{\ket{\scL,\downarrow},\ket{\scL,\uparrow},\ket{\scR,\downarrow},\ket{\scR,\uparrow} \}$, the associated matrix $v$ is of the form,
\begin{equation}
v = \begin{pmatrix}
0 & 0 & 0 & 1/2 \\
0 & 0 & 1/2 & 0 \\
0 & 1/2 & 0 & 0 \\
1/2 & 0 & 0 & 0
\end{pmatrix},
\end{equation}
When this matrix is diagonalized to yield $\bar{v}$, the diagonal elements of $\bar{v}$ represent $\bar{v}_k$ associated with Eq.~\ref{diagonalized}. Upon diagonalizing, we have that,
\begin{equation}
\bar{v} = \begin{pmatrix}
-1/2 & 0 & 0 & 0 \\
0 & -1/2 & 0 & 0 \\
0 & 0 & 1/2 & 0 \\
0 & 0 & 0 & 1/2
\end{pmatrix},
\end{equation}
and so, $\{\bar{v}_k\} = \{-1/2,-1/2,1/2,1/2 \}$. The states (which are the eigenvectors of $v$) attached to these coefficients are $\{ \ket{\bar{x}_k} \} = \frac{1}{\sqrt{2}} \{-\ket{\scL,\downarrow}+\ket{\scR,\uparrow} , -\ket{\scL,\uparrow}+\ket{\scR,\downarrow},\ket{\scL,\downarrow}+\ket{\scR,\uparrow} , \ket{\scL,\uparrow}+\ket{\scR,\downarrow} \}$. Plugging these coefficients and states into Eq.~\ref{diagonalized}, one finds $\ket{\Phi_x} $ represented in first quantization.  Because this expansion is unique~\cite{ItalianPRA} and there are four terms, the Schmidt rank is four, implying $\ket{\Phi_x} $ is particle entangled.

Even when interactions are absent, it is possible to create probabilistic particle entanglement through a cooperation of quantum statistics and measurement~\cite{Lester_measurement_2017}. This can occur when distinguishable particles -- either photons via their polarization or atoms via their spin -- are placed in an HOM interferometer. The role of measurement is to post-select on a particular symmetrization manifold of a distinguishable state, which allows for selection of a triplet ($\ket{\Phi_+} $) or singlet ($\ket{\Phi_-} $) in the distinguishing degree of freedom, both of which describe particle-entangled states. For example, consider a pair of bosonic atoms in opposing spin states undergoing a double-well beam-splitter interaction. Because the particles are distinguishable, they do not undergo HOM interference; half the time they will reside in the same well and half the time they will not. Due to the symmetrization of the quantum state, post-selecting on those experiments in which the atoms end up in different wells in turn post-selects on the spin state of the atoms being in a singlet.  Hence, measurement allows the probabilistic synthesis of a particle-entangled state. Indeed, this basic principle underlies linear quantum computer architectures with photons, whose lack of interaction does not permit the same kind of entangling gates seen with ions or neutral atoms.

Finally, even in the absence of a spin degree-of-freedom, it is possible to generate particle entanglement. In Sec.~\ref{sec:interaction} we undertake a detailed analysis of interactions of spinless bosons, i.e.~only particles with only a spatial degree of freedom, and we give an example of how the Schmidt rank can characterize particle entanglement even in this case (Sec.~\ref{sec:app_schmidt}).


\section{Interaction in two and many-particle interference}
\label{sec:interaction}
Photon experiments naturally achieve a noninteracting system where the dominant effect of quantum statistics is clear.  However, in ultracold atomic systems interatomic  interaction can easily be relevant. The interplay between quantum statistics and interactions can be highly non-trivial and give rise to novel phenomena absent in photonic systems, and of course is a natural part of condensed-matter systems. In this section we will focus on this interplay.
We start by considering the simplest possible system of two particles in a double well.   Next we discuss the situation of multiple atoms in a double well and finally treat the most interesting case consisting of many atoms in many wells. We consider both the case of bosonic and fermionic particles.  When analyzing the non-interacting case we of course arrive at analogous results to those considered for multi-particle interference with photons in Sec.~\ref{sec:manyparticlephotons}, but it is interesting to see this appear out of formalisms more typical to interacting cold atoms, such as the Bose-Hubbard model.

\subsection{Double well }

\subsubsection{Two atoms}

From a physical standpoint, the main role of interactions is to introduce energy shifts that depend on the number of particles per mode (site). These shifts can then suppress the HOM dip observed with non-interacting particles. This effect can be cleanly seen in the case of a double well potential with two bosonic particles.

The Hamiltonian for the interacting atoms is given by

\eq
\hat H_{\text{DW}} = - J ( \hat a^\dagger_1 \hat a_2 + \hat a^\dagger_2 \hat a_1  ) + \sum_{j=1}^2\frac{ U}{2}  \hat n_j  \left( \hat n_j -1\right),   \label{DW}
\en Here $U$ is the interaction energy cost of having two particles in the same lattice site and  $\hat n_j = \hat a^\dagger_j \hat a_j$ is the number operator. For two atoms  the Hilbert space is spanned by three states $\ket{2,0}$, $\ket{0,2}$, and $\ket{1,1}$.

A good starting point to study the HOM effect is the state
$\ket{\pm}=\frac{\ket{2,0}\pm \ket{0,2}}{\sqrt{2}}$. Due to the reflection symmetry of the Hamiltonian only the $\ket{+}$ state couples to the $\ket{1,1}$ state. The  $\ket{-}$ is decoupled from the rest.   The dynamics  in the subspace spanned by the states  $\ket{1,1}, \ket{+}$ reduces to

\eq
\hat H_{\text{DW}} = \left(
              \begin{array}{cc}
                0 & -2 J \\
                -2J & U \\
              \end{array}
            \right)
   \label{DWM}
\en Consequently if one prepares  the system in the $\ket{\psi(0)}=\ket{1,1}$ state at time $t=0$, the probability of remaining in that state after some time $t$, $P_{1,1}=|\langle 1,1\ket{\psi(t)} |^2$ is given by
\eq
P_{1,1}(t)=\left(1 - \frac{16 J^2}{ \Omega^2}\sin^2[\Omega t/2] \right)
\en with $\Omega^2= 16 J^2 + U^2$.   From this expression one concludes  that   $P_{1,1}(t)\geq U^2/(16 J^2 + U^2)$. The minimum value  ${\rm min}(P_{1,1}(t))=0$ is only reached for  noninteracting particles, $U=0$, at  the HOM dip when  $t_{HOM}=\pi/(4J)$. Any finite $U$ introduces an energy shift between singly and  doubly occupied states. This shift  introduces a net detuning that brings the system away from the resonant condition required for  perfect  destructive interference. In the strong interacting limit $U\gg J$ then  $P_{1,1}(t)\to 1$. The absence of the HOM dip reflects the suppression of tunneling due to a large onsite repulsion. Two identical fermions will also exhibit $P_{1,1}(t)= 1$, due to the  Pauli blockade. In fact, 1D bosons in the $U/J\gg 1$ limit, and with nearest-neighbor tunneling, exhibit the phenomena known as fermionization \cite{Girardeau1960}. In this case, any observable that can be written in terms of site occupation (number) operators  maps to the one computed if the particles were instead non-interacting fermions. Nevertheless, strongly interacting bosons still obey bosonic statistics.

 Fermionic particles  with an internal spin degree of freedom $\sigma=\uparrow, \downarrow$ that allows the population of sites with two particles in the same mode can exhibit similar behavior to bosonic particles.  For the relevant case where one of the atoms is $\uparrow$  and the other  $\downarrow$, the Hilbert space consists of the following states:
$\ket{\uparrow\downarrow,0}$,$\ket{0,\uparrow\downarrow}$,$\ket{\uparrow,\downarrow}$ and $\ket{\downarrow,\uparrow}$.  Here $\ket{\uparrow,\downarrow}$ and $\ket{\downarrow,\uparrow}$ are states with one atom per well in opposite spin configuration and $\ket{\uparrow\downarrow,0}$,$\ket{0,\uparrow\downarrow}$  states with two atoms in one well and zero in the other. The spins in the doubly occupied states  must  be in a singlet spin configuration due to fermionic statistics. The Hamiltonian is now

\eq
\hat H_{\text{FDW}} = - J \sum_\sigma ( {\hat c}^{\sigma \dagger}_1 {\hat  c}^\sigma_2 + {\hat c}^{\sigma\dagger}_2  {\hat c}^\sigma_1 ) + U  {\hat n}^\uparrow_j   {\hat n}^\downarrow_j ,   \label{DW}
\en with ${{\hat c}^\sigma}_i$  fermionic annihilation operators acting on a particle at site $i=1,2$ and spin $\sigma$. A good basis to use is  the one spanned by the states: $\ket{\pm}=(\ket{\uparrow \downarrow,0}\pm\ket{0,\uparrow\downarrow})/\sqrt{2}$, $\ket{S}=(\ket{\uparrow,\downarrow}-\ket{\downarrow,\uparrow})/\sqrt{2}$, and
$\ket{T}=(\ket{\uparrow,\downarrow}+\ket{\downarrow,\uparrow})/\sqrt{2}$.

Because tunneling preserves the spin of the particles, and the system exhibits reflection symmetry, the states $\ket{T}$ and $\ket{-} $ can not tunnel and are decoupled from other states. They have energy $0$ and $U$ respectively. The states  $\ket{S}$ and $\ket{+} $ on the other hand are coupled by tunneling and obey exactly the same equations of motions as those described by Eq. \ref{DWM}. The identical behavior between these states and the corresponding bosonic states can be easily understood by noticing that two fermions in a  spin singlet state, have the same symmetric spatial wave function as two identical bosons.

Now we proceed to discuss the case of many-atoms and many modes. When the atoms are non-interacting their behavior maps to the one we have already briefly described in Sec.~\ref{sec:manyparticlephotons}. Below, however, we will  use a slightly different but complementary approach   to describe the physics. Not only can it help the reader to gain a broader perspective,  but can also facilitate the understanding of interference  in the presence of interactions.

\subsubsection{Many atoms}

Let's now consider the case of many bosonic particles, $N$, in a double well.  To connect to the $N=2$ case let us first understand the
 non-interacting limit. For that let us remind ourselves of the Hadamard Lemma,
\eq
e^{-i \hat H_{\text{DW}} t} \hat a^\dagger_j e^{i \hat H_{\text{DW}} t}  = \sum_{m=0}^\infty \frac {(-i t)^m} {m!} [ \hat H_{\text{DW}}  , \hat a_j^\dagger ]_{m} ,
\en
where
\eq
[ X, Y ]_{m} =[ X, [ X, Y]_{m-1} ] ,~~~ [X, Y]_0 = Y
\en
Furthermore, we have
\eq
[ a, (a^\dagger)^N ] = N (a^{\dagger})^{N-1} \\
\left[ a^N, a^{\dagger} \right] = N a^{N-1}
\en
Using those properties for $U=0$ one finds
\eq
[ \hat H_{\text{DW}}, \hat a^\dagger_m ]_1 &=& J \hat a^\dagger_{3-m}  \label{Jeq} \\
e^{-i H_{\text{DW}} t} \hat a^\dagger_j e^{i \hat H_{\text{DW}} t} & = & \sum_{m=0}^\infty \frac {(-i t J)^m} {m!}  \hat a_{2-(j+1 \mod 2)}^\dagger  \\
&=& \cos( t J) \hat a_j^\dagger - i \sin ( t J) \hat a_{3-j}^\dagger  \label{timeevoU}
\en where we clearly see oscillations between the two wells.
By applying the time-evolution for a finite time $t=\pi/(4J)$, we obtain again a balanced beam-splitter in the double well.

To understand the role of interactions in the case of many bosonic particles, $N$, in a double well one can use the  Schwinger bosons representation. This naturally emerges by noting that the Hilbert space  is spanned by the Fock states  $\ket{n_1,N- n_1}$, with $0\geq n_1\geq N$, which correspond to having $n_1$ particles in one well and $N- n_1$ in the other.
 Those states can be mapped to a spin $S=N/2$ particle with magnetic quantum number  $S_z=n_1-N/2$. Using this mapping the Hamiltonian can be  written as
\eq
\hat H_{\text{DW}} = - 2J ( \hat S_x  ) + U \hat S_z^2  ,   \label{DWS}
\en where $\hat S_x=(\hat a^\dagger_1 \hat a_2 + \hat a^\dagger_2 \hat a_1 )/2$, $\hat S_y=(\hat a^\dagger_1 \hat a_2 - \hat a^\dagger_2 \hat a_1 )/(2i)$ and $\hat S_z=(\hat a^\dagger_1 \hat a_1 - \hat a^\dagger_2 \hat a_2 )/2$. In Eq.~\ref{DWS} we have omitted constant terms proportional to $(\hat{n}_1+\hat{n}_2)^2$ and $(\hat{n}_1+\hat{n}_2)$, since the Hamiltonian conserves particle number . This Hamiltonian is the so called Lipkin-Meshkov-Glick Model \cite{Vidal2004,Larson2010,Roman2008,Latorre2005,Homan2008,Cui2008}.
From Eq.~\ref{DWS}, it is clear that tunneling  generates rotations of the collective Bloch vector along the $x$ direction.

\begin{figure}
\includegraphics[width=.85\linewidth]{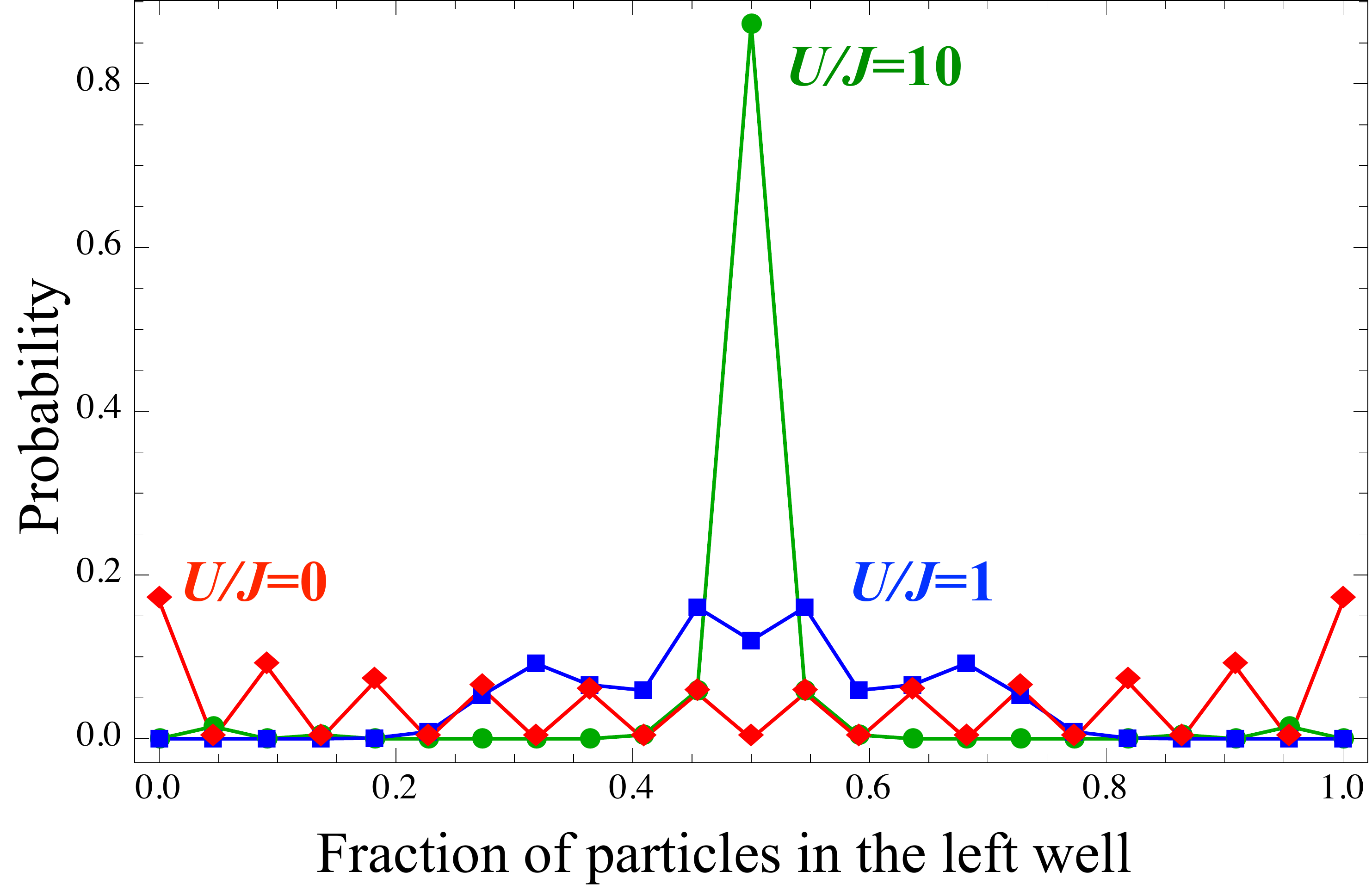}
\caption{Two-mode many-particle scattering example, including interactions between the particles.  This plot is to be compared to Fig.~\ref{fig:twomodes}, which uses the same parameters, but studies non-interacting particles. The interactions lead to a breakdown of the suppression odd-particle events, a reduction in even outcomes compared to the non-interacting case, and narrow the distribution so that the initial input state is preserved in the output.
}\label{fig:Int_DW}
\end{figure}

In the context of the HOM effect, the relevant observable is the  probability to populate a state with $n_1$ atoms in one of the wells and  $N-n_1$ in the other, $P_{n_1,N-n_1}(t)=|\langle n_1,N- n_1|\psi(t)\rangle|^2$. When  evaluated at $t_{HOM}$, $P_{n_1,N-n_1}(t_{HOM})$ is the probability for a particular fraction to be encountered in one of the two outputs of the beamsplitter as discussed in Sec.~\ref{sec:manyparticlephotons} and displayed in Fig.~\ref{fig:twomodes}. This probability can be computed from the moments of $\hat{S}_z^n(t)$. Because $\langle{\hat{S}_z^n(t)\rangle}= \sum_{m=0}^N (A)_{n m} P_{m,N-m}$
 with $A_{nm}=(m-N/2)^n$, the later equation can be inverted and we obtain

\eq
P_{m,N-m}(t)=\sum_{n=0}^N (A^{-1})_{mn}\langle{\hat{S}_z^n(t)\rangle} \label{mop},
\en \noindent where $A^{-1} $ is the inverse of the matrix $A$. For the non-interacting system $\hat{S}_z(t)= \cos(2 J t) \hat{S}_z(0)+ \sin(2 J t)   \hat{S}_y(0)$, and therefore
\eq
\langle{ \hat{S}_z^n(t)\rangle}=\left\langle\left(\cos(2 J t) \hat{S}_z(0)+ \sin(2 J t)   \hat{S}_y(0)
\right)^n \right \rangle  \label{mom}
\en   One clearly sees that $\langle{ \hat{S}_z^n(t)\rangle}$ exhibits interference fringes with periodicity $\tau=\pi/J$ in agrement with Eq. \ref{timeevoU}. Moreover at $t_{HOM}$, $\langle \hat{S}_z^{n}(t_{HOM})\rangle=\langle \hat{S}_y^{n}(0)\rangle$.    If the initial state is an eigenstate of $S_z(0)$   then  all odd moments  of $\hat{S}_y(0)$
vanish  and only the even ones remain finite. If at $t=0$,    $P_{N/2,N/2}(0)=1$  at $t_{HOM}$
 $P_{N/2,N/2}(t_{HOM})=0$ reflecting destructive interference. For example for the  $N=2$ case  discussed before,  $P_{1,1}(t_{HOM})=1-\langle \hat{S}_z^2(t_{HOM})\rangle= 1-\langle \hat{S}_y^2(0)\rangle=0$, consistent with the expectations. In Fig.\ref{fig:Int_DW} we show   $P_{11,11}(t_{HOM})$ for the non-interacting system which is the double well analog of the beamsplitter case shown in Fig. \ref{fig:twomodes}.

As in the $N=2$ case,  a finite $U$ introduces an energy cost to  move one   particle from one well to the other disrupting the perfect destructive interference exhibited by non-interacting particles, i.e.~$P_{N/2,N/2}(t_{HOM})>0$.
 In the large $U$ limit the particle motion is  suppressed and  $\lim_{U\to \infty }P_{m,N-m}(t)\to  P_{m,N-m}(0)$ (see Fig.~\ref{fig:Int_DW}) similar to the $N=2$ case.  From this analysis we see that although the many-particle system exhibits a more complex behavior, one  can identify similar features seen in the two particle case.

\subsection{Bose-Hubbard: N atoms in L lattice sites }

Let's now consider the situation where $N$ atoms are prepared in $L$ sites of a deep optical lattice where  to an excellent approximation only nearest-neighbour tunneling is relevant. We assume that interactions are  local and experienced when two or more atoms occupy the same lattice site. The Hamiltonian describing this situation is the Bose-Hubbard model given by

\eq
\hat H_{\text{BH}} = - J \sum_{\langle{i,j\rangle}} ( \hat a^\dagger_i \hat a_j  ) +  \frac{ U}{2} \sum_{j=1}^N  \hat n_j  \left( \hat n_j -1\right), \label{DW} \en    Here $\langle{i,j\rangle}$ means nearest-neighbor lattice sites.

The   Bose-Hubbard model dynamics  shares many common features to the ones described in the double well setup. To illustrate  the similarities  we will first consider two limiting cases, $U=0$ and $U\to \infty$. We will restrict the analysis to a 1D lattice.

\subsubsection{$U=0$}
In the non-interacting limit,  the Hamiltonian reduces to $\hat H_{\text{BH}} = - J \sum_{\langle{i,j\rangle}} ( \hat a^\dagger_i \hat a_j  )$ and    can be exactly diagonalized by the  unitary transformation (assuming periodic boundary conditions) $\hat b_q =\frac{1}{\sqrt{L}}\sum_{j=1}^{L} e^{iqj} \hat a_j$. Here $q= \frac{2 \pi}{L} n_q$ is the quasimomentum $n_q=0,\dots, L-1$.
Using it, one can compute the evolution of the operator

\eq
\hat a^\dagger_n (t)=e^{-i H_{\text{BH}} t} \hat a^\dagger_n e^{i \hat H_{\text{BH}} t} = \sum_{m=0}^L A_m^n(t) \hat a^\dagger_m(0) \label{sin}\en here $A_m^n(t)=i^{m-n}\mathcal{J}_{m-n}( 2J t)$, with $\mathcal{J}_n(x)$  Bessel functions of the first kind. Note that in contrast to the double-well case the dynamics is no longer periodic with $J$. Instead the particle exhibits a quantum random walk  as a function of time. Using the fact that for $x\gg n^2$  $\mathcal{J}_{n}( x)\sim \sqrt{\frac{2}{\pi x}} \cos(x-\pi/4-n \pi/2)$, one can see that transport of the particle is ballistic, because the probability density of a particle initially localized at site $n$ is  $\langle\hat n_m (t) \rangle=\langle\hat a^\dagger_m (t) \hat a_m (t)\rangle=|\mathcal{J}_{m-n}( 2J t)|^2$, which expands linearly in time.

In the case of $N$ identical bosons initially localized at sites $\{n_1,\dots,n_N\}$, the evolution of the equal time normal ordered  correlator $P_{m_1,\dots,m_N}(t)=\langle \hat a^\dagger_{m_1}(t) \hat a^\dagger_{m_2}(t)\cdots \hat a^\dagger_{m_N}(t) \hat a_{m_1}(t) \hat a_{m_2}(t)\cdots  \hat a_{m_N}(t)\rangle$
is  given by the permanent (see also Sec.~\ref{sec:manyparticlephotons}) of the following matrix
\begin{eqnarray}
&&P_{m_1,\dots,m_N}(t) = \left |{\rm Perm} ( A)  \right|^2 \notag\\
&&A =\left(
                                                 \begin{array}{cccc}
                                                  A_{m_1}^{n_1}(t) & A_{m_2}^{n_1}(t) & \dots & A_{m_N}^{n_1}(t)  \\
                                                   A_{m_1}^{n_2}(t) & A_{m_2}^{n_2}(t) & \dots & A_{m_N}^{n_2}(t)  \\
                                                   \cdot & \cdot & \cdot & \cdot \\
                                                   A_{m_1}^{n_N}(t) & A_{m_2}^{n_N}(t) & \cdot & A_{m_N}^{n_N}(t)  \\
                                                 \end{array}\right)
\end{eqnarray}.

This observable displays the key ingredients required for the HOM: (1) quantum statistics, which manifests in the requirement to compute the permanent of single particle functions, and, (2) coherent superposition of paths, which arise because of the nonzero contribution of the interference terms when the square of the permanent is computed. To explicitly observe the contribution from (1) and (2), let's consider the simple case of two particles initially located at sites $L/2$ and $L/2+1$.  The probability of finding the particle at the same initial location is  given by $P_{n_1=L/2,n_2=L/2+1}(t)=  | A_{n_1}^{n_1}(t) A_{n_2}^{n_2}(t)+A_{n_1}^{n_2}(t) A_{n_2}^{n_1}(t)|^2=|\mathcal{J}_{0}^2( 2J t)-\mathcal{J}_{1}^2( 2J t)|^2\sim \sin[4 t J]^2/(\pi^2 (t J)^2)$. At $t_{HOM}$, $P_{L/2,L/2+1}(t_{HOM})\sim 0$,
 which can be interpreted as the equivalent destructive interference effect that gives rise to the  HOM dip for $L=2$, as well as a consequence of symmetrization (encapsulated in the  permanent) and a  non-zero interference term $ A_{n_1}^{n_1}(t) A_{n_2}^{n_2}(t)A_{n_1}^{n_2}(t) A_{n_2}^{n_1}(t)$.  A similar picture holds for many particles.

\subsubsection{$U\to \infty$}
In the strongly interacting limit, the repulsion between the particles forbids the population of sites with more than one particle and in one dimension, the local observables of the bosonic system resemble the ones of non-interacting fermions. In this limit, instead of the permanent one needs to compute  a determinant \eq \small{
P_{m_1,\dots,m_N}(t) =  \left |{\rm det}  \left(
                                                 A
                                               \right)\right|^2}
\en  If one  analyzes the two particle case, as we did above, one obtains

$P_{n_1=L/2,n_2=L/2+1}(t)=  | A_{n_1}^{n_1}(t) A_{n_2}^{n_2}(t)-A_{n_2}^{n_1}(t) A_{n_1}^{n_2}(t)|^2=|\mathcal{J}_{0}^2( 2J t)+\mathcal{J}_{1}^2( 2J t)|^2\sim 1/(\pi^2 (t J)^2)$. At $t_{HOM}$, a  constructive interference arises due to the different sign of the term $- A_{n_1}^{n_1}(t) A_{n_2}^{n_2}(t)A_{n_1}^{n_2}(t) A_{n_2}^{n_1}(t)$ in striking contrast to the destructive one found for the $U=0$ limit.

\subsubsection{Finite $U$}

For the intermediate interacting regime the dynamics is more complex because finite $U$ introduces  additional terms  in Eq.~\ref{sin} of the form

\begin{eqnarray}
e^{-i H_{\text{BH}} t} \hat a^\dagger_n e^{i \hat H_{\text{BH}} t} &=& \sum_{m} A_m^n(t) \hat a^\dagger_m \notag \\
&&+  U \sum_{m} M_{m} (\hat a^\dagger_m)^2 \hat a_m + \text{etc.} \label{BHG}
\end{eqnarray}  The above equation implies that the set of states that can be produced with both the tunneling and interactions is larger
than what can be produced with just tunneling \cite{TichyEntanglementBosonsFermions}.  During the dynamics tunneling and interactions  couple the modes and the particles, respectively~\cite{Preiss2015} and thus  mode and particle entangled states are not mutually exclusive.  The tunneling produces the number fluctuations among the modes (in this case, these are the lattice sites), but the interactions allow for states that are truly particle entangled. The states exhibit number correlations inaccessible to systems exhibiting only tunneling. To further elucidate this notion, in the next section we discuss a specific example of a particle-entangled state of spinless, indistinguishable bosons. 


 \subsubsection{Application of Schmidt rank to Bose-Hubbard example}
 \label{sec:app_schmidt}

In Sec.~\ref{ref:Schmidt}, we showed how the Schmidt rank indicates entanglement for the canonical example of a Bell-State of distinguishable particles. The dynamics of the Bose-Hubbard model illustrate how particle-entangled states of spinless bosons --- that is, indistinguishable particles with only a spatial degree of freedom --- can emerge as well. For example, in the recent publication of Ref.~\cite{Preiss2015}, a pair of strongly-interacting particles undergoes a quantum walk under the influence of the Bose-Hubbard Hamiltonian, with $J\ll U$.  In this limit the two particles form a  bound state which tunnels as a single object with  an effective tunneling rate $J^2/U$ . Starting from the initial state $\ket{\psi(0)} = (a_0^\dagger)^2\ket 0 $, the particles remain largely bound and undergo a quantum walk with an effective tunneling rate of $2J^2/U$. In time, the state evolves as,
 \begin{equation}
\vert \psi(t) \rangle = \sum_n i^n \mathcal{J}_n(\frac{4 J^2}{U} t)\frac{(a_n^\dagger)^2}{\sqrt{2}} \ket 0 ,
 \end{equation}
where $n$ is the site index, $\mathcal{J}_n$ is the Bessel function of the first kind, and $t$ is time. For times where more than two terms in the summation are non-vanishing, the state exhibits particle entanglement. This can be seen by rewriting $\vert \psi(t)\rangle$ in the first quantization  decomposition of Eq.~\ref{diagonalized},
\begin{equation}
\ket{\psi(t)_{fq}} =\sum_n i^n \mathcal{J}_n(\frac{4 J^2}{U} t) \ket{n}_1\ket{n}_2,
\end{equation}
where $\ket n$ is the localized single-particle wavefunction of site $n$. This summation will exhibit more than two terms whenever the doublon is delocalized over more than two sites, which occurs quickly after a few effective tunneling times of $(J^2/U)^{-1}$.  By contrast, in the absence of interactions, two particles would independently tunnel leading to a symmetrized wavefunction of Schmidt rank equal to 1.


\section{Entanglement entropy}

In addition to the challenges of classifying and generating quantum entanglement, its quantification presents yet another experimental obstacle. While HOM interference can be exploited to generate varying types of entanglement, generalized HOM interference can be harnessed to detect entanglement in many-body states,
as we discuss in the following.

\begin{figure*}
\includegraphics[scale=.9]{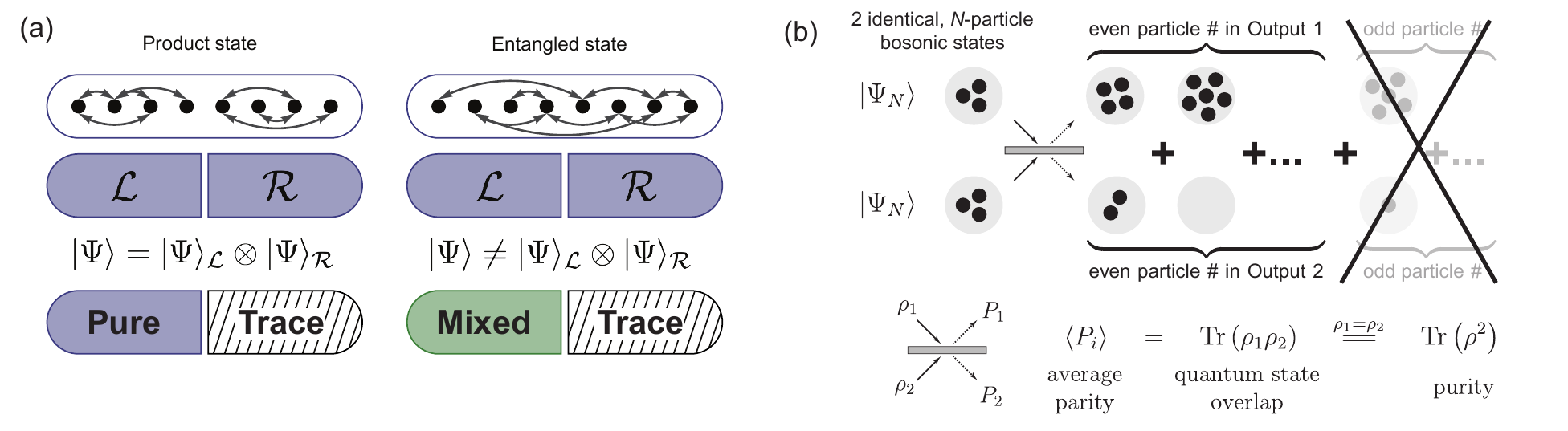}
\caption{Entanglement entropy. (a) A quantum state can be partitioned into subsystems $\scL$ and $\scR$, that represent subspaces of the full Hilbert space in which the quantum state lives. When the quantum state is pure and non-entangled, it can be written as a separable product of pure states in  $\scL$ and $\scR$. When it is pure and entangled, the full quantum state cannot be written this way, and in either subspace the state is mixed, i.e. entropic. (b)  Through the use of beamsplitter interactions and number-resolved counting, it is possible to measure local and global quantum state purities. In particular, the counting statistics (the parity) after the application of a beamsplitter is sensitive to quantum state overlap of interfered states, from which purities can be deduced. This allows for the measurement of entanglement entropy. Images adapted from Ref.~\cite{Islam2015}.
\label{fig:ee_figure}}
\end{figure*}

One way of mathematically quantifying entanglement is through the entanglement entropy. It quantifies the degree of non-separability across a cut that partitions the system into subsystems. For example, for a chain of spins, one might ask how the spins in the right half of the chain are entangled with those in the left half. Or, for a Bose-Hubbard chain, we might consider mode entanglement between the right and the left half of the system. In general, we can ask whether a state is separable by introducing a cut through the system, thereby defining two subsystems each with an associated complete basis in the left and right Hilbert spaces ($\mathcal{H_L}$ and $\mathcal{H_R}$) (Fig.~\ref{fig:ee_figure}).

For any  bipartite state $\ket{\psi}$, there is a unique decomposition --- the Schmidt decomposition --- such that,
\begin{equation}
\ket{\psi}=\sum_j c_j \ket{\chi^\scL_j}\ket{\chi^\scR_j}\,
\label{eq:schmidt}
\end{equation}
where for the different terms in the sum we have $\langle \chi^\scL_{i}  \ket{\chi^\scL_j} = \langle  \chi^\scR_{i} \ket{\chi^\scR_j} = \delta_{ij}$, and $\ket{\chi^{\scL,\scR}} \in \mathcal{H_{L,R}}$. The state $\ket{\psi}$ is non-separable whenever there is more than one non-vanishing  term in this sum; otherwise, the state is a product state with respect to the partitioning. Entanglement entropy is linked to the degree of mixing of the reduced density matrix of just one of the subsystems, namely,
\begin{equation}
\label{eq:red_state}
\rho_{\scS}=\sum_j \vert c_j \vert^2 \ket{\chi^{\scS}_j}\bra{\chi^{\scS}_j}\,
\end{equation}
where $\scS=\scL$, $\scR$.
Whenever $\ket{\psi}$ is entangled, $\rho_{\scS}$ is a mixed state because there is more than one term in the sum in Eq.~\ref{eq:red_state}.
When a subsystem state is mixed and the full system state is pure, the full system state is entangled.
Hence, we refer to the loss of local purity, or, equivalently, the generation of local entropy, as entanglement entropy. It can be quantified with the von-Neumann entanglement entropy $S_{vN} = \mathrm{Tr}(\rho_{\scL} \mathrm{log}(\rho_{\scL}))= \mathrm{Tr}(\rho_{\scR} \mathrm{log}(\rho_{\scR}))$ or the R\'enyi entropies $S_n = \mathrm{log}(\mathrm{Tr}(\rho_{\scL}^n))= \mathrm{log}(\mathrm{Tr}(\rho_{\scR}^n))$. The $n=2$ R\'enyi entropy is the logarithm of the purity of the subsystem density matrix.

By definition, entanglement is invariant under coherent dynamics within the individual subsystems (local unitaries) \cite{horodecki:865};
that is, a state ${\cal U}_\scL\otimes{\cal U}_\scR\ket{\psi}$, with unitaries ${\cal U}_\scL$ and ${\cal U}_\scR$ on the left and right subsystem,
carries the same amount of entanglement as the state $\ket{\psi}$.
The experimental assessment of entanglement following a setup with both left and right subsystems as input of a beam splitter \cite{PhysRevLett.91.087903,PhysRevA.83.042318} will not necessarily assign the same amount of entanglement to $\ket{\psi}$ and to ${\cal U}_\scL\otimes{\cal U}_\scR\ket{\psi}$,
 but there are functions of quantum states that are invariant under all local unitaries.
The norm is a particularly simple example, but it provides no information on entanglement properties.
All other invariant functions are necessarily non-linear in $\ket{\psi}$ and $\bra{\psi}$ (or the density matrix $\varrho$ in the case of mixed states), and the simplest non-linear function is bi-linear.
Because all measurements on quantum mechanical systems yield expectation values $\bra{\psi}A\ket{\psi}$ of an observable $A$, i.e.~a linear function, non-linear functions can be measured only with extra effort.

A way to realize the measurement of bilinear functions is to work with two versions, also called twins, of a quantum system simultaneously.
The two twins of the left half are described in terms of the Hilbert space ${\mathcal H}_\scL\otimes {\mathcal H}_\scL$ rather than the Hilbert space ${\mathcal H}_\scL$ of a single twin.
Analogously, the two twins of the right half are described in terms of the Hilbert space ${\mathcal H}_\scR\otimes {\mathcal H}_\scR$.
The central idea is to prepare the two twin systems, each consisting of a left and a right twin, in the same fashion \cite{brun,PhysRevLett.108.110503}.
The state of both twins together is then given by $\ket{\psi}_1\otimes\ket{\psi}_2$, where $\ket{\psi}$ is the state of an individual twin, and the index `$1$' and `$2$' labels the two twins.
Expectation values of such twin states provide a characterization and quantification of entanglement, which has the desired invariance properties \cite{PhysRevLett.90.167901,PhysRevLett.95.240407,horodecki:052323,PhysRevLett.88.217901,PhysRevA.86.052330}.

The degree of mixing of reduced density matrices $\hat \varrho_{\scL,\scR}$ that quantifies the entanglement of the underlying pure state \cite{meyerwallach,brennen,Bennett2015,mintert:260502}
can be expressed as
\begin{equation}
\tr \hat \varrho_{\scS}^2=\tr\ \Pi_{\scS}\ \left(\hat \varrho_{\scS}\otimes \hat \varrho_{\scS}\right)\
\end{equation}
in terms of two identical, reduced density matrices $\varrho_{\scS}$, where $\Pi_{\scS}$ is the permutation operator defined via the relation $\Pi_{\scS}\ket{i}_{\scS_1}\otimes\ket{j}_{\scS_2}=\ket{j}_{\scS_1}\otimes\ket{i}_{\scS_2}$,
where $\ket{i/j}_{\scS_1}$ are subsystem states of the first twin of subsystem $\scS$ and $\ket{i/j}_{\scS_2}$ are states of the second twin.
Consequently, proper entanglement measures for pure states $\ket{\psi}$ can be constructed based on expectation values of the permutation operators $\Pi_{\scL}$ and $\Pi_{\scR}$ with respect to the twin-state $\ket{\psi}\otimes\ket{\psi}$ \cite{demkowicz-dobrzanski:052303,PhysRep}.

In the case of mixed states, the local purities alone can not assess entanglement properties,
 because for example a direct product of mixed single sub-system states would be consistent with low local purities.
The entanglement of mixed states can, however, be estimated if in addition to local purities also the purity of the entire state is accessible \cite{PhysRevLett.98.140505,PhysRevA.78.022308}.
The resulting quantitative estimate is particularly good for weakly mixed states \cite{mintert:012336};
and it remains valid under imperfect preparation of the copies \cite{mintert:052302} as long as they remain uncorrelated \cite{PhysRevLett.102.190503}.

Such assessments of entanglement in terms of purities were implemented for photons
\cite{Walborn:2006lr,walborn:032338,schmid:260505}
but current experiments with atomic gases permit to implement analog measurements for substantially larger systems \cite{PhysRevA.72.042335,PhysRevLett.93.110501,PhysRevLett.109.020505}.

The realization of the required measurement of permutation symmetry for the twin-states of each subsystem
can indeed be based on a beam splitter with the action
\begin{equation}
\begin{split}
\hat a_{\scS_1}^\dagger&\to(\hat a_{\scS_1}^\dagger+\hat a_{\scS_2}^\dagger)/\sqrt{2}\\
\hat a_{\scS_2}^\dagger&\to( \hat a_{\scS_1}^\dagger-\hat a_{\scS_2}^\dagger)/\sqrt{2}\ ,
\label{eq:bs}
\end{split}
\end{equation}
where $\hat a_{\scS_1}^\dagger$ and $\hat a_{\scS_2}^\dagger$ are the creation operators for the two twin-subsystems under consideration.
Its action on a Fock state
\begin{equation}
\ket{ij}_{\scS}=\ket{i}_{\scS_1}\otimes\ket{j}_{\scS_2}=\frac{1}{\sqrt{i!j!}}(\hat a_{\scS_1}^\dagger)^i(\hat a_{\scS_2}^\dagger)^j\ket{0}
\end{equation}
reads
\begin{equation}
\ket{ij}_{\scS}\to\frac{(\hat a_{\scS_1}^\dagger+\hat a_{\scS_2}^\dagger)^i(\hat a_{\scS_1}^\dagger-\hat a_{\scS_2}^\dagger)^j}{2^{i+j}\sqrt{i!j!}}\ket{0}=\ket{\chi_{ij}}_{\scS}.
\label{eq:perm}
\end{equation}
Because the application of the permutation $\Pi$ amounts to the exchange of $\hat a_{\scS_1}^\dagger$ and $\hat a_{\scS_2}^\dagger$, from Eq.~\eqref{eq:perm} one can directly read off the relation
\eq
\Pi_{\scS} \ket{\chi_{ij}}_{\scS}=(-1)^j\ket{\chi_{ij}}_{\scS}.
\en
That is, the beamsplitter maps states with an even number of particles in the second twin-subsystem onto states the are symmetric under $\Pi$, whereas states with an odd number of particles in the second twin-subsystem are mapped onto anti-symmetric states.
Because the beamsplitter operation is self-inverse, also any symmetric state is mapped onto a state with an even number of particles in the second twin-subsystem, and any anti-symmetric state is mapped onto a state with an odd particle number.
Consequently, local purities can be measured with the help of the beamsplitter in Eq.~\eqref{eq:bs} and a measurement of particle number that can be realized with state of the art optical read-out methods \cite{Sherson:2010ys,nature:7269}.

For the measurement of the global purity of an $N$-body system one merely needs to notice that the global permutation is just the tensor product of all local permutations. That is, the ability to measure all local purities is sufficient to measure 
the purity of the $N$-body states, and of the purities for all $M$-body ($M<N$) reduced density matrices \cite{aolita:050501,PhysRevLett.109.020505}.

In a recent experiment \cite{Islam2015} with $^{87}$Rb atoms in an optical lattice entanglement for systems of up to four atoms (per twin) was observed.
The system can be modeled well with a Bose-Hubbard Hamiltonian (Eq.\eqref{DW}).
It induces the beam splitter relation defined in Eq.~\eqref{timeevo},
that differs from Eq.~\eqref{eq:bs} in terms of the phase shift that a particle acquires when tunneling.
The difference between both beam splitters could be compensated through appropriate phase shifts $\exp(\pm i\frac{\pi}{2} \hat a_{\scS_2}^\dagger \hat a_{\scS_2})$ before and after the beam splitter.
The measurement of the particle number is un-affected by any phase shift after the second beam splitter, so that no phase shift after the beam splitter needs to be applied.
If, however, the necessary phase shift before the beam-splitter is not applied,
a measurement of particle number in the second twin systems, results in the measurement of $\tr \hat \varrho\exp(i\frac{\pi}{2} \hat a_{\scS_2}^\dagger \hat a_{\scS_2})\hat \varrho\exp(-i\frac{\pi}{2} \hat a_{\scS_2}^\dagger \hat a_{\scS_2})$ rather than $\tr \hat \varrho^2$.
Due to the Cauchy-Schwarz inequality $(\tr \hat \varrho_1 \hat \varrho_2)^2\le\tr \hat \varrho_1^2\ \tr \hat \varrho_2^2$,
this still results in a reliable lower bound of the purity.
In addition, for Fock states, the phase shift before the beam splitter results in a global phase without observable consequences, so that the beam splitter Eq.~\eqref{timeevo} including phase shift results in the correct assessment of purity for all experimentally investigated states.


In the regime of strong interactions $U\gg J$, the ground state is a Fock state with no mode entanglement,
but with growing $J/U$, the atoms get increasingly delocalized and mode entanglement builds up.
Rather high purities for single-site and two-site reduced density matrices were observed for $U/J\gtrsim 10$,
and, in the regime $10\gtrsim U/J\gtrsim 1$ a rapid decrease in purity was observed, consistent with the theoretical expectation.
The purity of the four-atom state showed no significant change in purity during the variation of $U/J$, which underlines that the changes in local purities are indeed a result of mode entanglement, and not merely of an increase of entropy that could result from experimental imperfections.

In the superfluid regime $J\gg U$, one expects that the dynamics following the preparation of a Fock state would result in periodic oscillations of mode entanglement.
For a system with two atoms and two sites, the dynamics following the initial state $\ket{1,1}$,
would result in a balanced superposition of all three accessible Fock states $\ket{0,2}$, $\ket{1,1}$ and $\ket{2,0}$ (at $Jt=\arctan(\sqrt{2})/(2\pi)) \sim 0.68$, i.e.~the state with highest mode entanglement achievable in this system.
At $Jt=\pi/(4)$ one would expect a balanced superposition of the two states $\ket{0,2}$ and $\ket{2,0}$ which has less mode entanglement.
Subsequently the system evolves again into a maximally entangled state, followed by the separable initial state~\cite{Islam2015}.

\section{Concluding Remarks}

Increasingly, ultracold atom experiments have turned to single-atom imaging and manipulation to gain a new level of understanding.  With this perspective,  a variety of experiments akin to those with single photons have appeared, and the Hong-Ou-Mandel effect with atoms is exemplary of such newly-enabled experiments. A goal of this article was to discuss fundamental concepts that consequently arise -- such as many-particle interference and entanglement of identical particles -- which until now were not prominent players in ultracold atom studies. Bringing these concepts to the fore also emphasizes the new opportunities that arise with ultracold atoms, due to efficient detection, the ability to introduce interactions, and by being massive. Applying these capabilities to experimental directions inspired by photonic protocols have, and will, continue to lead to new directions.

Future experiments may harness many-particle interference of atoms in a variety of ways.  As addressed above, the incorporation of quantum beamsplitters into cold atom technology has allowed measurement of entanglement entropy through many-particle interference, yielding an intersection of ideas in quantum optics, atomic physics, and condensed matter physics. Many-particle interference is also fundamental to the boson sampling problem, for which cold atom technologies could augment the accessible system sizes and complexity. Furthermore, photons are routinely entangled in polarization or frequency degrees of freedom by utilizing measurement and quantum statistics; the entanglement results from multiparticle interference, and circumvents the challenge of producing entanglement with non-interacting particles. 

From a theoretical standpoint, the subtleties surrounding entanglement of identical particles raises several questions. This article addressed the case of entanglement between two particles, relying on the CSOP prescription. As just noted, there are many future pursuits that entail large samples of atoms; delineating the various forms of entanglement in such cases will be of significant import, and are not within the scope of the CSOP criterion. Relatedly, experiments are naturally imperfect and create only approximately pure quantum states, that is, the states are mixed: theoretically defining particle entanglement in mixed quantum states of identical particles will be similarly important, both for classifying and devising methods for quantifying entanglement.
%
%
%

\acknowledgements{We thank Alain Aspect, Michael Foss-Feig, Markus Greiner, Murray Holland, Brian Lester, Yiheng Lin, and Christoph Westbrook for conversations.  We acknowledge funding from the NSF under grant number PHYS 1734006, the David and Lucile Packard Foundation, the Office of Naval Research under award number N00014-17-1-2245, AFOSR FA9550-13-1-0086, AFOSR-MURI Advanced
Quantum Materials, NIST and DARPA W911NF-16-1-0576 through ARO.}
\clearpage


\begin{thebibliography}{100}

\bibitem{HOM}
C.~K. Hong, Z.~Y. Ou, and L. Mandel, {\em Measurement of subpicosecond time
  intervals between two photons by interference}, Phys. Rev. Lett. {\bf 59},
  2044  (1987).

\bibitem{Knill2001}
E. Knill, R. Laflamme, and G.~J. Milburn, {\em A scheme for efficient quantum
  computation with linear optics}, Nature {\bf 409},  46  (2001).

\bibitem{Aaronson:2010fk}
S. Aaronson and A. Arkhipov, arXiv:1011.3245 (unpublished).

\bibitem{Walraff2013}
C. Lang {\it et~al.}, {\em Correlations, indistinguishability and entanglement
  in Hong-Ou-Mandel experiments at microwave frequencies}, Nat. Phys. {\bf 9},
  345  (2013).

\bibitem{Toyoda2015}
K. Toyoda, R. Hiji, A. Noguchi, and S. Urabe, {\em Hong--Ou--Mandel
  interference of two phonons in trapped ions}, Nature {\bf 527},  74  (2015).

\bibitem{electron}
E. Bocquillon {\it et~al.}, {\em Coherence and Indistinguishability of Single
  Electrons Emitted by Independent Sources}, Science {\bf 339},  1054  (2013).

\bibitem{Li2016}
J. Li {\it et~al.}, {\em Hong-Ou-Mandel Interference between Two Deterministic
  Collective Excitations in an Atomic Ensemble}, Phys. Rev. Lett. {\bf 117},
  180501  (2016).

\bibitem{Kaufman2014}
A.~M. Kaufman {\it et~al.}, {\em Two-particle quantum interference in
  tunnel-coupled optical tweezers}, Science {\bf 345},  306  (2014).

\bibitem{Bakr2011}
W.~S. Bakr {\it et~al.}, {\em Orbital excitation blockade and algorithmic
  cooling in quantum gases}, Nature {\bf 480},  500  (2011).

\bibitem{Preiss2015}
P.~M. Preiss {\it et~al.}, {\em Strongly correlated quantum walks in optical
  lattices}, Science {\bf 347},  1229  (2015).

\bibitem{Lopes2015}
R. Lopes {\it et~al.}, {\em Atomic Hong-Ou-Mandel experiment}, Nature {\bf
  520},  66  (2015).

\bibitem{Roos2017}
C.~F. Roos, A. Alberti, D. Meschede, P. Hauke, and H. Haeffner, {\em Revealing
  quantum statistics with a pair of distant atoms}, Phys. Rev. Lett.  160401
  (2017).

\bibitem{TichyThesis}
M.~C. Tichy, Ph.D. thesis, Physikalisches Institut, 2011.

\bibitem{ItalianPRA}
G. Ghirardi and L. Marinatto, {\em General criterion for the entanglement of
  two indistinguishable particles}, Phys. Rev. A {\bf 70},  012109  (2004).

\bibitem{TichyJOP}
M.~C. Tichy, F. Mintert, and A. Buchleitner, {\em Essential entanglement for
  atomic and molecular physics}, Journal of Physics B: Atomic, Molecular and
  Optical Physics {\bf 44},  192001  (2011).

\bibitem{Cirac2012}
J.~I. Cirac, {\em Entanglement in many-body quantum systems}, arXiv:1205.3742v1
   (2012).

\bibitem{Barnum2004}
H. Barnum, E. Knill, G. Ortiz, R. Somma, and L. Viola, {\em A
  Subsystem-Independent Generalization of Entanglement}, Phys. Rev. Lett. {\bf
  92},  107902  (2004).

\bibitem{Paunkovic2004}
N. Paunkovic, Ph.D. thesis, Oxford, 2004.

\bibitem{Lim:2005qt}
Y.~L. Lim and A. Beige, {\em Generalized Hong-Ou-Mandel experiments with bosons
  and fermions}, N. J. Phys. {\bf 7},  155  (2005).

\bibitem{Tichy:2012NJP}
M.~C. Tichy, M. Tiersch, F. Mintert, and A. Buchleitner, {\em Many-particle
  interference beyond many-boson and many-fermion statistics}, New J. Phys {\bf
  14},  093015  (2012).

\bibitem{PhysRevA.91.013811}
A. Crespi, {\em Suppression laws for multiparticle interference in Sylvester
  interferometers}, Phys. Rev. A {\bf 91},  013811  (2015).

\bibitem{Ou:2007ly}
Z.~Y.~J. Ou, {\em Multi-Photon Quantum Interference} (Springer, New York,
  2007).

\bibitem{Spagnolo:2013fk}
N. Spagnolo {\it et~al.}, {\em Three-photon bosonic coalescence in an
  integrated tritter}, Nat. Commun. {\bf 4},  1606  (2013).

\bibitem{Ra:2013kx}
Y.-S. Ra {\it et~al.}, {\em Nonmonotonic quantum-to-classical transition in
  multiparticle interference}, Proc. Natl. Acad. Sci. USA {\bf 110},  1227
  (2013).

\bibitem{Jin2016}
R.-B. Jin {\it et~al.}, {\em Detection-dependent six-photon Holland-Burnett
  state interference},  {\bf 6},  36914 EP   (2016).

\bibitem{Huang2011}
Y.-F. Huang {\it et~al.}, {\em Experimental generation of an eight-photon
  Greenberger--Horne--Zeilinger state},  {\bf 2},  546 EP   (2011).

\bibitem{Yao2012}
X.-C. Yao {\it et~al.}, {\em Observation of eight-photon entanglement}, Nat
  Photon {\bf 6},  225  (2012).

\bibitem{PhysRevLett.117.210502}
X.-L. Wang {\it et~al.}, {\em Experimental Ten-Photon Entanglement}, Phys. Rev.
  Lett. {\bf 117},  210502  (2016).

\bibitem{Daley}
A.~J. Daley, H. Pichler, J. Schachenmayer, and P. Zoller, {\em Measuring
  Entanglement Growth in Quench Dynamics of Bosons in an Optical Lattice},
  Phys. Rev. Lett. {\bf 109},  020505  (2012).

\bibitem{Islam2015}
R. Islam {\it et~al.}, {\em Measuring entanglement entropy in a quantum
  many-body system}, Nature {\bf 528},  77  (2015).

\bibitem{Kaufman2016}
A.~M. Kaufman {\it et~al.}, {\em Quantum thermalization through entanglement in
  an isolated many-body system}, Science {\bf 353},  794  (2016).

\bibitem{Obrien2013}
J.~C.~F. Matthews {\it et~al.}, {\em Observing fermionic statistics with
  photons in arbitrary processes}, Sci. Rep. {\bf 3},  1539  (2013).

\bibitem{Keilmann2010}
T. Keilmann, S. Lanzmich, I. McCulloch, and M. Roncaglia, {\em Statistically
  induced Phase Transitions: Turning Bosons smoothly via Anyons into Fermions},
  Nat. Comm. {\bf 2},  361  (2010).

\bibitem{Rarity1996}
J.~G. Rarity, P.~R. Tapster, and R. Loudon,  in {\em Quantum Interferometry},
  edited by F. DeMartini, G. Denardo, and Y. Shih (VCH Weinheim, ADDRESS,
  1996), p.\ 211.

\bibitem{Riedmatten2003}
H. de~Riedmatten, I. Marcikic, W. Tittel, H. Zbinden, and N. Gisin, {\em
  Quantum interference with photon pairs created in spatially separated
  sources}, Phys. Rev. A {\bf 67},  022301  (2003).

\bibitem{Kaltenbach2006}
R. Kaltenbaek, B. Blauensteiner, M. Zukowski, M. Aspelmeyer, and A. Zeilinger,
  {\em Experiment interference of independent photons}, Phys. Rev. Lett. {\bf
  96},    (2006).

\bibitem{Pittman1996}
{\em Can two-photon interference be considered the interference of two
  photons}, Phys. Rev. Lett. {\bf 77},  1917  (1996).

\bibitem{Beugnon2006}
J. Beugnon {\it et~al.}, {\em Quantum interference between two single photons
  emitted by independently trapped atoms}, Nature {\bf 440},  779  (2006).

\bibitem{Kaufman2012}
A.~M. Kaufman, B.~J. Lester, and C.~A. Regal, {\em Cooling a Single Atom in an
  Optical Tweezer to Its Quantum Ground State}, Phys. Rev. X {\bf 2},  041014
  (2012).

\bibitem{Lopesthesis}
R. Lopes, Ph.D. thesis, Institut D'Optique, 2015.

\bibitem{Anderson1995}
M.~H. Anderson, J.~R. Ensher, M.~R. Matthews, C.~E. Wieman, and E.~A. Cornell,
  {\em Observation of Bose-Einstein Condensation in a Dilute Atomic Vapor},
  Science {\bf 269},  198  (1995).

\bibitem{Andrews1997}
M.~R. Andrews {\it et~al.}, {\em Observation of Interference Between Two Bose
  Condensates}, Science {\bf 275},  637  (1997).

\bibitem{Ottl2005}
A. Ottl, S. Ritter, M. Kohl, and T. Esslinger, {\em Correlations and Counting
  Statistics of an Atom Laser}, Phys. Rev. Lett. {\bf 95},  090404  (2005).

\bibitem{Jeltes2007}
T. Jeltes {\it et~al.}, {\em Comparison of the Hanbury Brown--Twiss effect for
  bosons and fermions}, Nature {\bf 445},  402  (2007).

\bibitem{Dall2010}
R. Dall {\it et~al.}, {\em observation of atomic speckle and Hanbury
  Brown--Twiss correlations in guided matter waves}, Nature Comm. {\bf 2},  291
   (2011).

\bibitem{Schellekens2005}
M. Schellekens {\it et~al.}, {\em Hanbury Brown Twiss effect for ultracold
  quantum gases}, Science {\bf 310},  648  (2005).

\bibitem{Perrin2012}
A. Perrin {\it et~al.}, {\em Hanbury Brown and Twiss correlations across the
  Bose-Einstein condensation threshold}, Nat. Phys. {\bf 8},  195  (2012).

\bibitem{BlochQGM}
J.~F. Sherson {\it et~al.}, {\em Single-atom-resolved fluorescence imaging of
  an atomic Mott insulator}, Nature {\bf 467},  68  (2010).

\bibitem{Simon2011}
J. Simon {\it et~al.}, {\em Quantum simulation of antiferromagnetic spin chains
  in an optical lattice}, Nature {\bf 472},  307  (2011).

\bibitem{Kaufmanthesis}
A. Kaufman, Ph.D. thesis, University of Colorado, 2015.

\bibitem{Strabley}
J. Sebby-Strabley {\it et~al.}, {\em {Preparing and probing atomic number
  states with an atom interferometer}}, Phys. Rev. Lett. {\bf 98},  200405
  (2007).

\bibitem{MeschedePNAS}
A. Steffen {\it et~al.}, {\em Digital atom interferometer with single particle
  control on a discretized space-time geometry}, Proceedings of the National
  Academy of Sciences {\bf 109},  9770  (2012).

\bibitem{Robens2016}
C. Robens, S. Brakhane, D. Meschede, and A. Alberti,  in {\em Laser
  Spectroscopy Proceedings of the XXII International Conference} (PUBLISHER,
  ADDRESS, 2015), p.\ 1.

\bibitem{Greiner2002}
M. Greiner, O. Mandel, T. Esslinger, T.~W. Hansch, and I. Bloch, {\em Quantum
  phase transition from a superfluid to a Mott insulator in a gas of ultracold
  atoms}, Nature {\bf 415},  39  (2002).

\bibitem{Altman2004}
E. Altman, E. Demler, and M.~D. Lukin, {\em Probing many-body states of
  ultracold atoms via noise correlations}, Phys. Rev. A {\bf 70},  013603
  (2004).

\bibitem{Greiner2005}
M. Greiner, C. Regal, J. Stewart, and D. Jin, {\em Spatial quantum noise
  interferometry in expanding ultracold atom clouds}, Phys. Rev. Lett. {\bf
  94},  110401  (2005).

\bibitem{Folling2005}
S. F\"{o}lling {\it et~al.}, {\em Spatial quantum noise interferometry in
  expanding ultracold atom clouds}, Nature {\bf 434},  481  (2005).

\bibitem{Glauber1963b}
R.~J. Glauber, {\em Photon Correlations}, Phys. Rev. Lett. {\bf 10},  84
  (1963).

\bibitem{Scully}
M.~O. Scully and M.~S. Zubairy, {\em Quantum Optics} (Cambridge University
  Press, ADDRESS, 1997).

\bibitem{Meystre}
P. Meystre and M. Sargent, {\em Elements of Quantum Optics} (Springer, Berlin,
  2007).

\bibitem{Loudon}
R. Loudon, {\em The Quantum Theory of Light} (Oxford University Press, Oxford,
  2000).

\bibitem{Dussarrat2017}
P. Dussarrat {\it et~al.}, {\em A two-particle, four-mode interferometer for
  atoms}, arXiv:1707.01279v1  (2017).

\bibitem{Brunner2017}
T. Br\"unner, G. Dufour, A. Rodriguez, and A. Buchleitner, {\em Signatures of
  indistinguishability in bosonic many-body dynamics}, arXiv:1710.08876
  (2017).

\bibitem{Dufour2017}
G. Dufour {\it et~al.}, {\em Many-particle interference in a two-component
  bosonic Josephson junction: an all-optical simulation}, New Journal of
  Physics {\bf 19},  125015  (2017).

\bibitem{Mullin:2006fk}
W.~J. Mullin, R. Krotkov, and F. Laloe, {\em The origin of the phase in the
  interference of Bose-Einstein condensates}, Am. J. Phys. {\bf 74},  880
  (2006).

\bibitem{Laloe:2010uq}
F. Lalo\"e and W. Mullin, {\em Quantum properties of a single beam splitter},
  Found. Phys. {\bf 42},  53  (2012).

\bibitem{Campos:1989fk}
R.~A. Campos, B.~E.~A. Saleh, and M.~C. Teich, {\em Quantum-mechanical lossless
  beam splitter: SU(2) symmetry and photon statistics}, Phys. Rev. A {\bf 40},
  1371  (1989).

\bibitem{nockdipm}
M. Nock, Master's thesis, Universit\"at Konstanz, 2006.

\bibitem{younsikraNatComm}
Y.-S. Ra {\it et~al.}, {\em Observation of detection-dependent multi-photon
  coherence times}, Nat. Commun. {\bf 4},  2451  (2013).

\bibitem{Reck:1994zp}
M. Reck, A. Zeilinger, H.~J. Bernstein, and P. Bertani, {\em Experimental
  Realization of Any Discrete Unitary Operator}, Phys. Rev. Lett. {\bf 73},  58
   (1994).

\bibitem{Scheel:2004hc}
S. Scheel, arxiv:0406127 quant-ph (unpublished).

\bibitem{Spring15022013}
J.~B. Spring {\it et~al.}, {\em Boson Sampling on a Photonic Chip}, Science
  {\bf 339},  798  (2013).

\bibitem{Tillmann:2012ys}
M. Tillmann {\it et~al.}, {\em Experimental Boson Sampling}, Nat. Photon. {\bf
  7},  540  (2013).

\bibitem{Crespi:2012vn}
A. Crespi {\it et~al.}, {\em Experimental boson sampling in arbitrary
  integrated photonic circuits}, Nat. Photon. {\bf 7},  545  (2013).

\bibitem{Broome15022013}
M.~A. Broome {\it et~al.}, {\em Photonic Boson Sampling in a Tunable Circuit},
  Science {\bf 339},  794  (2013).

\bibitem{Tichy:2010ZT}
M.~C. Tichy, M. Tiersch, F. de~Melo, F. Mintert, and A. Buchleitner, {\em
  Zero-Transmission Law for Multiport Beam Splitters}, Phys. Rev. Lett. {\bf
  104},  220405  (2010).

\bibitem{Tichy:2013lq}
M.~C. Tichy, K. Mayer, A. Buchleitner, and K. M{\o}lmer, {\em Stringent and
  efficient assessment of Boson-Sampling devices}, Phys. Rev. Lett. {\bf 113},
  020502  (2014).

\bibitem{Dittel2016}
C. Dittel, R. Keil, and G. Weihs, arXiv:1607.00836 (unpublished).

\bibitem{Crespi2016}
A. Crespi {\it et~al.}, {\em Suppression law of quantum states in a 3D photonic
  fast Fourier transform chip}, Nat. Commun. {\bf 7},  10469  (2016).

\bibitem{Spagnolo:2013eu}
N. Spagnolo {\it et~al.}, {\em Efficient experimental validation of photonic
  boson sampling against the uniform distribution}, Nat. Photon. {\bf 8},  615
  (2014).

\bibitem{Javorsek:2000bh}
D. Javorsek {\it et~al.}, {\em New Experimental Test of the Pauli Exclusion
  Principle Using Accelerator Mass Spectrometry}, Phys. Rev. Lett. {\bf 85},
  2701  (2000).

\bibitem{Collaboration:2010ly}
{Borexino Collaboration}, {\em New experimental limits on the Pauli-forbidden
  transitions in\^{}{\{}12{\}}C nuclei obtained with 485 days Borexino data},
  Phys. Rev. C {\bf 81},  034317  (2010).

\bibitem{PhysRevLett.111.130503}
N. Spagnolo {\it et~al.}, {\em General Rules for Bosonic Bunching in Multimode
  Interferometers}, Phys. Rev. Lett. {\bf 111},  130503  (2013).

\bibitem{TichyMultiDimPerm}
M.~C. Tichy, {\em Sampling of partially distinguishable bosons and the relation
  to the multi-dimensional permanent}, Phys. Rev. A {\bf 91},  022316  (2015).

\bibitem{Dirac:1930vn}
P.~A.~M. Dirac, {\em The Principles of Quantum Mechanics (Fourth edition,
  1958)} (Oxford University Press, Oxford, 1930).

\bibitem{PhysRevLett.108.010502}
L. Sansoni {\it et~al.}, {\em Two-Particle Bosonic-Fermionic Quantum Walk via
  Integrated Photonics}, Phys. Rev. Lett. {\bf 108},  010502  (2012).

\bibitem{Gard:13}
B.~T. Gard, R.~M. Cross, P.~M. Anisimov, H. Lee, and J.~P. Dowling, {\em
  Quantum random walks with multiphoton interference and high-order correlation
  functions}, J. Opt. Soc. Am. B {\bf 30},  1538  (2013).

\bibitem{natphys:345}
C. Lang {\it et~al.}, {\em Correlations, indistinguishability and entanglement
  in Hong-Ou-Mandel experiments at microwave frequencies}, Nat Phys {\bf 9},
  345  (2013).

\bibitem{Ray2011}
M.~R. Ray and S.~J. van Enk, {\em Verifying entanglement in the Hong-Ou-Mandel
  dip}, Phys. Rev. A {\bf 83},  042318  (2011).

\bibitem{Sackett2000}
C.~A. Sackett {\it et~al.}, {\em Experimental entanglement of four particles},
  Nature {\bf 404},  256  (2000).

\bibitem{Lester_measurement_2017}
B.~J. Lester {\it et~al.}, {\em Measurement-based entanglement of
  non-interacting bosonic atoms}, arXiv:1712.06633  (2017).

\bibitem{Girardeau1960}
M. Girardeau, {\em Relationship between Systems of Impenetrable Bosons and
  Fermions in One Dimension}, Journal of Mathematical Physics {\bf 1},  516
  (1960).

\bibitem{Vidal2004}
J. Vidal, G. Palacios, and C. Aslangul, {\em Entanglement dynamics in the
  Lipkin-Meshkov-Glick model}, Phys. Rev. A {\bf 70},  062304  (2004).

\bibitem{Larson2010}
J. Larson, {\em Circuit QED scheme for the realization of the
  Lipkin-Meshkov-Glick model}, EPL (Europhysics Letters) {\bf 90},  54001
  (2010).

\bibitem{Roman2008}
R. Or\'us, S. Dusuel, and J. Vidal, {\em Equivalence of Critical Scaling Laws
  for Many-Body Entanglement in the Lipkin-Meshkov-Glick Model}, Phys. Rev.
  Lett. {\bf 101},  025701  (2008).

\bibitem{Latorre2005}
J.~I. Latorre, R. Or\'us, E. Rico, and J. Vidal, {\em Entanglement entropy in
  the Lipkin-Meshkov-Glick model}, Phys. Rev. A {\bf 71},  064101  (2005).

\bibitem{Homan2008}
H.-M. Kwok, W.-Q. Ning, S.-J. Gu, and H.-Q. Lin, {\em Quantum criticality of
  the Lipkin-Meshkov-Glick model in terms of fidelity susceptibility}, Phys.
  Rev. E {\bf 78},  032103  (2008).

\bibitem{Cui2008}
H.~T. Cui, {\em Multiparticle entanglement in the Lipkin-Meshkov-Glick model},
  Phys. Rev. A {\bf 77},  052105  (2008).

\bibitem{TichyEntanglementBosonsFermions}
M.~C. Tichy, F. Mintert, and A. Buchleitner, {\em Limits to multipartite
  entanglement generation with bosons and fermions}, Phys. Rev. A {\bf 87},
  022319  (2013).

\bibitem{horodecki:865}
R. Horodecki, P. Horodecki, M. Horodecki, and K. Horodecki, {\em Quantum
  entanglement}, Rev. Mod. Phys. {\bf 81},    (2009).

\bibitem{PhysRevLett.91.087903}
G. Burkard and D. Loss, {\em Lower Bound for Electron Spin Entanglement from
  Beam Splitter Current Correlations}, Phys. Rev. Lett. {\bf 91},  087903
  (2003).

\bibitem{PhysRevA.83.042318}
M.~R. Ray and S.~J. van Enk, {\em Verifying entanglement in the Hong-Ou-Mandel
  dip}, Phys. Rev. A {\bf 83},  042318  (2011).

\bibitem{brun}
T.~A. Brun, {\em Measuring polynomial functions of states}, Quant. Inf. Comp.
  {\bf 4},  401  (2004).

\bibitem{PhysRevLett.108.110503}
S.~J. van Enk and C.~W.~J. Beenakker, {\em Measuring
  $\mathrm{Tr}{\ensuremath{\rho}}^{n}$ on Single Copies of $\ensuremath{\rho}$
  Using Random Measurements}, Phys. Rev. Lett. {\bf 108},  110503  (2012).

\bibitem{PhysRevLett.90.167901}
P. Horodecki, {\em Measuring Quantum Entanglement without Prior State
  Reconstruction}, Phys. Rev. Lett. {\bf 90},  167901  (2003).

\bibitem{PhysRevLett.95.240407}
F.~A. Bovino {\it et~al.}, {\em Direct Measurement of Nonlinear Properties of
  Bipartite Quantum States}, Phys. Rev. Lett. {\bf 95},  240407  (2005).

\bibitem{horodecki:052323}
P. Horodecki, R. Augusiak, and M. Demianowicz, {\em General construction of
  noiseless networks detecting entanglement with the help of linear maps},
  Phys. Rev. A {\bf 74},  052323  (2006).

\bibitem{PhysRevLett.88.217901}
A.~K. Ekert {\it et~al.}, {\em Direct Estimations of Linear and Nonlinear
  Functionals of a Quantum State}, Phys. Rev. Lett. {\bf 88},  217901  (2002).

\bibitem{PhysRevA.86.052330}
F. Mintert, B. Salwey, and A. Buchleitner, {\em Many-body entanglement:
  Permutations and equivalence classes}, Phys. Rev. A {\bf 86},  052330
  (2012).

\bibitem{meyerwallach}
D. Meyer and N. Wallach, , J. of Math. Phys {\bf 43},  4273  (2002).

\bibitem{brennen}
G.~K. Brennen, {\em An observable measure of entanglement for pure states of
  multi-qubit systems}, Quant. Inf. Comp. {\bf 3},  619  (2003).

\bibitem{Bennett2015}
C.~H. Bennett, D.~P. DiVincenzo, J.~A. Smolin, and W.~K. Wootters, {\em
  Mixed-state entanglement and quantum error correction}, Phys. Rev. A {\bf
  54},  3824  (1996).

\bibitem{mintert:260502}
F. Mintert, M. Ku\'{s}, and A. Buchleitner, {\em Concurrence of Mixed
  Multipartite Quantum States}, Phys. Rev. Lett. {\bf 95},  260502  (2005).

\bibitem{demkowicz-dobrzanski:052303}
R. Demkowicz-Dobrza\'{n}ski, A. Buchleitner, M. Ku\'{s}, and F. Mintert, {\em
  Evaluable multipartite entanglement measures: Multipartite concurrences as
  entanglement monotones}, Phys. Rev. A {\bf 74},  052303  (2006).

\bibitem{PhysRep}
F. Mintert, A.~R.~R. Carvalho, M. Ku\'{s}, and A. Buchleitner, {\em Measures
  and dynamics of entangled states}, Phys. Rep. {\bf 415},  207  (2005).

\bibitem{PhysRevLett.98.140505}
F. Mintert and A. Buchleitner, {\em Observable Entanglement Measure for Mixed
  Quantum States}, Phys. Rev. Lett. {\bf 98},  140505  (2007).

\bibitem{PhysRevA.78.022308}
L. Aolita, A. Buchleitner, and F. Mintert, {\em Scalable method to estimate
  experimentally the entanglement of multipartite systems}, Phys. Rev. A {\bf
  78},  022308  (2008).

\bibitem{mintert:012336}
F. Mintert and A. Buchleitner, {\em Concurrence of quasipure quantum states},
  Phys. Rev. A {\bf 72},  012336  (2005).

\bibitem{mintert:052302}
F. Mintert, {\em Concurrence via entanglement witnesses}, Phys. Rev. A {\bf
  75},  052302  (2007).

\bibitem{PhysRevLett.102.190503}
S.~J. van Enk, {\em Direct Measurements of Entanglement and Permutation
  Symmetry}, Phys. Rev. Lett. {\bf 102},  190503  (2009).

\bibitem{Walborn:2006lr}
S.~P. Walborn, P.~H. Souto~Ribeiro, L. Davidovich, F. Mintert, and A.
  Buchleitner, {\em Experimental determination of entanglement with a single
  measurement}, Nature {\bf 440},  1022  (2006).

\bibitem{walborn:032338}
S.~P. Walborn, P.~H.~S. Ribeiro, L. Davidovich, F. Mintert, and A. Buchleitner,
  {\em Experimental determination of entanglement by a projective measurement},
  Phys. Rev. A {\bf 75},  032338  (2007).

\bibitem{schmid:260505}
C. Schmid {\it et~al.}, {\em Experimental Direct Observation of Mixed State
  Entanglement}, Phys. Rev. Lett. {\bf 101},  260505  (2008).

\bibitem{PhysRevA.72.042335}
R.~N. Palmer, C. Moura~Alves, and D. Jaksch, {\em Detection and
  characterization of multipartite entanglement in optical lattices}, Phys.
  Rev. A {\bf 72},  042335  (2005).

\bibitem{PhysRevLett.93.110501}
C. Moura~Alves and D. Jaksch, {\em Multipartite Entanglement Detection in
  Bosons}, Phys. Rev. Lett. {\bf 93},  110501  (2004).

\bibitem{PhysRevLett.109.020505}
A.~J. Daley, H. Pichler, J. Schachenmayer, and P. Zoller, {\em Measuring
  Entanglement Growth in Quench Dynamics of Bosons in an Optical Lattice},
  Phys. Rev. Lett. {\bf 109},  020505  (2012).

\bibitem{Sherson:2010ys}
J.~F. Sherson {\it et~al.}, {\em Single-atom-resolved fluorescence imaging of
  an atomic Mott insulator}, Nature {\bf 467},  68  (2010).

\bibitem{nature:7269}
W.~S. Bakr, J.~I. Gillen, A. Peng, S. Folling, and M. Greiner, {\em A quantum
  gas microscope for detecting single atoms in a Hubbard-regime optical
  lattice}, Nature {\bf 462},  74  (2009).

\bibitem{aolita:050501}
L. Aolita and F. Mintert, {\em Measuring Multipartite Concurrence with a Single
  Factorizable Observable}, Phys. Rev. Lett. {\bf 97},  050501  (2006).

\end{thebibliography}

\end{document}